\pdfoutput=1
\documentclass[aps,amsfonts,amsmath,prd,preprint,nofootinbib]{revtex4}
\newcommand{\beq}{\begin{equation}}
\newcommand{\eeq}{\end{equation}}

\usepackage{graphicx,subfigure}

\begin{document}

\title{Accidental Inflation in the Landscape}

\author{Jose J. Blanco-Pillado${}^{a}$, Marta Gomez-Reino${}^{ b}$ and Konstantinos Metallinos${}^{a}$}
\affiliation{$^a$ Institute of Cosmology, 
Department of Physics and Astronomy\\ Tufts University, Medford, MA, US}
\affiliation{$^b$ Department of Physics, University of Oviedo, Oviedo, Spain.}

\def\changenote#1{\footnote{\bf #1}}

\begin{abstract}

We study some aspects of fine tuning in inflationary scenarios within 
string theory flux compactifications and, in particular,
in models of accidental inflation. We investigate the possibility that the apparent
fine-tuning of the low energy parameters of the theory needed to have inflation
can be generically obtained by scanning the values of the
fluxes over the landscape. Furthermore, we find that the existence 
of a landscape of eternal inflation in this model provides us with a natural theory of 
initial conditions for the inflationary period in our vacuum.
We demonstrate how these two effects work in a small corner of the landscape 
associated with the complex structure of the Calabi-Yau manifold
$P^4_{[1,1,1,6,9]}$ by numerically investigating the flux vacua of a reduced
moduli space. This allows us to obtain the distribution of observable 
parameters for inflation in this mini-landscape directly from 
the fluxes.

\end{abstract}

\maketitle
\thispagestyle{empty}
\section{Introduction}
\setcounter{page}{1}

The idea that our universe has undergone a period of 
inflation seems to fit the current observations of the cosmic microwave 
background \cite{WMAP7}. However, these observations have not
taught us, so far, much about its fundamental origin. It is
therefore interesting to analyze the different scenarios
that can give rise to inflation within a fundamental theory
with the hope that one could identify a distinctive signature
that allow us to understand the relevant dynamics that
were in play in these early stages of cosmic history.

These ideas have led many people to investigate the 
cosmological consequences of different high energy models and,
in particular, string theory (See, for example, some of the reviews on the
subject in \cite{String-Cosmology}). One problem
that has plagued this field from the beginning was the 
necessity of a model of stable compactification that reduces the
original $10d$ theory to its $4d$ low energy form. Recently 
the old Kaluza-Klein idea of flux compactifications \cite{FR} has been revived within
the string theory set up \cite{flux-compactifications}. These models
indicate the existence of perturbatively
stable vacua of the theory where we could find ourselves
today \cite{GKP,KKLT}. Nevertheless it is pretty clear by now that 
the mechanism of compactification is very far from unique,
and it seems very likely that this method of compactifying
would lead to a immense landscape of distinct vacua
\cite{BP,Susskind}. 
Some of these vacua would be similar to our four dimensional
universe, but others would have very different properties,
for example the value of the cosmological constant \cite{BP}, the 
low energy physics \cite{Lust} or even the number of large dimensions \cite{BPSPV-2}. 
Transitions between these vacua are allowed
by a tunneling event where a bubble of the new vacuum is
produced in the background of the parent vacuum. This 
process can continue due to the presence of metastable
vacua with positive cosmological constant leading
to the picture of an eternally inflating spacetime
where all vacua are explored in what has been collectively called 
the multiverse.

These ideas in models of flux compactification
inspired a flurry of papers on new scenarios of inflation
within string theory. Most of these studies concentrate
on identifying a particular sector of the effective
potential for the $4d$ fields that allows for inflation. 
We can classify these models depending whether the inflaton 
field is related to the position of a D-brane along the extra
dimensions (D-brane inflation \cite{D-Brane-Inflation}) or whether 
it parametrizes the shape and size of the internal
manifold (Modular Inflation \cite{Modular-Inflation}).

Modular inflation is a natural idea in any 
higher dimensional extension of the standard model since
the potential is already present in the construction to fix the moduli fields.
On the other hand, it is clear that this could be a very complicated 
function with many possible forms. In models with fluxes, this effective potential 
encodes the information about the quantized fluxes
that thread the cycles in the internal space. Turning on these
fluxes generates a change in energy as a function of the size of the
cycles, which in turn is interpreted as potential for these fields. Studying 
the implications of a particular set of fluxes implies that one 
is focusing on a single realization of the corresponding potential in the
landscape. This has been the approach used 
in most of the concrete models of inflation in string theory so far. 
This is a reasonable thing to do since cosmological 
observations would only allow us to see the last 
$60$ e-folds of inflation\footnote{This is not entirely true since
it is possible that a tunnelling transition between different
vacua could leave its
imprint in some of the cosmological observables that we can detect,
provided that the total amount of inflation within our bubble is 
not too large (See \cite{Open-Inflation} or more recently 
\cite{FKRMS,Open-Inflation-in-the-Landscape}). We will discuss this issue in
more detail later on in the paper.} which, presumably, would
happen within the effective potential with the same
set of fluxes. 

Furthermore, it is likely that there are many different
regions of the landscape that allow for inflation (even within the
same model) and 
if this is the case, we will have to face the question 
of what is the most likely inflationary scenario on the
string theory landscape. The answer to this question 
will require us to adopt some measure on how to give
probabilities to all these inflationary trajectories. This
is a hard problem that has been extensively investigated in the
last few years and although some progress seems to have emerged from these
studies there is not a clear consensus in the literature on how
to assign these probabilities (See \cite{Measure-Problem} for some
recent discussion on the subject). We will not have anything new to 
say here about the measure problem and concentrate on another issues where the
existence of a landscape can play a significant role, namely, in the fine tuning of the 
potential as well as the initial conditions required for a successful inflation
to occur. This is a first necessary step towards extracting
observational predictions in the landscape which would require
information about the underlying theory as well as the measure
problem \cite{Guth-Nomura}.

Most models of inflation so far
studied in string theory suffer from some kind of fine tuning
that, from a purely effective field theory perspective,
could be considered quite severe. However, changing the
fluxes along the internal manifold will in practice
allow the parameters in the effective potential to 
scan over different values so there could be many different
sets of fluxes that would lead to a particular value for the
coefficient in the potential compatible with observations. The most 
important example of this
phenomenon is, of course, the idea that the cosmological constant
problem could be solved by the vast numbers of possible vacua that
we have in the landscape, so some of them could land on the very
narrow observationally allowed region \cite{BP}. In this paper we will
show that this same idea could help with the 
fine tuning required to flatten the potential during 
inflation. 

There are several interesting
papers in the literature that try to model the complexity of the 
$4d$ effective potential in the landscape by statistical arguments, see for 
example 
\cite{Dynamical-fine-tuning,Inflation-in-the-Landscape,D-brane-inflection-2,D-brane-inflection-3}. 
In the following, we will compare our methods to some of these other 
approaches when appropriate.

The outline of the paper is as follows. We first provide, in Section II,
an overview of the methods that lead to the effective potential
in models of Type IIB flux compactifications. In Section III we show
how a simple example of fine-tuned accidental inflation \cite{AI} could arise in a
toy model for the landscape. In Section IV we explore a 
small corner of the landscape and find how the relevant parameters
of the model are scanned in this mini-landscape and we describe how a 
natural choice of initial conditions can arise in this models.
In Section V we give the distribution of the different
observational parameters obtained in our example of the landscape.
We conclude in Section VI with some comments and a general outlook.

\section{Review of Flux compactifications}

Most models of inflation based on the evolution of moduli fields
use, as a starting point, the low energy description of the
supersymmetric string theories, namely a supergravity theory. In our
case we will focus on Type IIB compactifications to four dimensions
on a Calabi-Yau (CY) orientifold, which reduces the scalar field theory to an
${\cal N}=1$ supergravity action for the moduli fields. In the following we 
will briefly describe the ingredients necessary to specify
the ${\cal N}=1$ theory for our moduli fields.

\subsection{{\cal N} =1 SUGRA actions}

The effective action of ${\cal N} =1$ supergravity with $n$ chiral fields, $\Phi_i$, 
will be fixed, for the purposes of this paper, once one specifies two different functions: a
holomorphic superpotential $W(\Phi_i)$,
and a real Kahler potential $K(\Phi_i, \Phi^\dagger_i)$. In terms of these functions the F-term 
scalar potential is given by\footnote{We fix $M_P=1$ and use the
standard notation $F_i=\frac{\partial }{\partial \Phi_i}$, 
$F_{\bar j}=\frac{\partial }{\partial \Phi^\dagger_{\bar j}}$,
$\cdots$, with $F$ being any function of the fields. Also note that
indices are lowered and raised with the Kahler metric $K_{i \bar j}$ and its inverse $K^{i \bar j}$ .}:
\begin{equation}
V_F= e^K \left(\sum_i{K^{i \bar j} D_i W 
D_{\bar j} {\bar W}- 3{{|W|^2}}}\right)
\end{equation}
where 
\begin{equation}
D_i W = \partial_i W+ \partial_i K W\,\,\,\,;\,\,\,\,D_{\bar
  j} 
\bar W = \partial_{\bar j} \bar W+ \partial_{\bar j} K \bar W\,.
\end{equation}
and the kinetic terms of the fields are computed from the Kahler
potential using,
\begin{equation}
{\cal L}_{kin} = K_{I{\bar J}} \partial_{\mu} \Phi^I \partial^{\mu}
\Phi^{\bar J}~.
\end{equation}
There may be, in general, other terms in the effective potential for the fields
that could be included in a supergravity action, but this simple
description would be sufficient for the type of models that we will describe below.

\subsubsection{Kinetic terms for the moduli}

One can obtain the information for the kinetic terms for
the moduli fields by performing a dimensional reduction of the $10d$
theory where the internal manifold is a Calabi-Yau
threefold \cite{Moduli-CY}. Following this procedure one identifies 
three different type of fields, the complex structure moduli, the Kahler moduli and
the axion-dilaton field. 

In the rest of this section we will briefly describe each of these
sectors of the scalar field space and their kinetic terms.

\begin{itemize}

\vspace{.5cm}
{\item Complex Structure}
\end{itemize}

\vspace{.5cm}
The Kahler potential for the complex structure 
moduli can be calculated to be \cite{Moduli-CY},
\begin{equation}
K_{cs} = - \log \left(i \int_M{\Omega \wedge \bar{\Omega}}\right) 
\end{equation}
where $\Omega$ denotes the holomorphic three form and
the integral is performed over the Calabi-Yau threefold $M$. We can
also rewrite the expression above as,
\begin{equation}
\label{Kcs}
K_{cs} = -\log~(- i \Pi^{\dagger} \cdot \Sigma \cdot \Pi)
\end{equation}
where we have introduced the simplectic matrix,
$\Sigma$, given by,
\begin{equation}
\label{Sigma}
\Sigma = \left(
\begin{array}{cc}
0 & {\bf 1} \\
{\bf -1} & 0 
\end{array}\right)
\end{equation}
as well as the period vector,
\begin{equation}
\Pi(w) = ( w^a, F_a)
\end{equation}
whose components are defined in terms of integrals of the holomorphic
three-form $\Omega$ over the tree-cycles, ${A^a, B_a}$, namely,
\begin{equation}
w^a = \int_{A^a}{\Omega}~~~~~~~ F_a = \int_{B_a}{\Omega}~. 
\end{equation}
On the other hand, $F_a$ can be computed in terms of the so-called 
prepotential $F$ by the relation, $F_a = \partial_a F(w)$. 
The complex structure moduli fields $z_i$ are then obtained via 
the normalization condition, $z_i = {\frac{w^i}{w^0}}$ so, having the
expression for the prepotential in terms of the $w^i$ coordinates one can write 
the Kahler potential as a function of the complex structure fields, 
$K_{cs}(z_i)$.

We will later describe in detail the complex structure moduli space of a particular
CY where all these definitions will become more clear.

\begin{itemize}
\vspace{.5cm}
{\item Kahler moduli}
\end{itemize}

\vspace{.5cm}
The Kahler potential for the Kahler moduli is given by,
\begin{equation}
K_K = - 2 \log ({\cal V})~,
\end{equation}
where ${\cal V}$ denotes the volume of the Calabi-Yau
in string units and can be written as,
\begin{equation}
{\cal V} = \int_M{J^3} = {1\over 6} \kappa_{jk} t^i t^j t^k
\end{equation}
where $J$ is the Kahler form of the CY and the $t^i$ fields 
denote the sizes of its 2-cycles. We can also, for later convenience
rewrite everything in terms of the volumes of the four-cycles,
$\tau_i$, by the following expression,
\begin{equation}
\tau_i = \partial_{t^i} {\cal V} = {1\over 2} \kappa_{ijk} t^j t^k.
\end{equation}
So far we have only discussed four dimensional massless
fields whose origin was a deformation of the internal 
geometry. But it is clear that there will be more massless
scalar fields coming from the compactification of different
p-forms present in the ten dimensional theory. In fact,
one can show that there are the same number of axionic
fields coming from the compactification of the Ramond-Ramond
4-form that pair up with the Kahler moduli fields to create
the complexified Kahler moduli, 
\begin{equation}
T_i = \tau_i + i \theta_i~,
\end{equation}
which are the fields that we will include in the ${\cal N}= 1$ 
supergravity description of the four dimensional theory. In this
paper we will be mainly interested in a simple toy model of a 
single Kahler moduli with Kahler function,
\begin{equation}
\label{Kahler-moduli-Kahler}
K_K = -3 \log \left(T + \bar{T}\right)~.
\end{equation}

\begin{itemize}
\vspace{.5cm}
{\item Dilaton}
\end{itemize}

\vspace{.5cm}
Finally, there is another important component of the 
four dimensional theory which is given by a complex field
composed of the zero mode of the dilaton ($\phi$) and the zero
mode of the axion field (the Ramond-Ramond zero form, $C_0$) 
already present in ten dimensions, namely,
\begin{equation}
\tau = C_0 + i e^{-\phi}~,
\end{equation}
such that in the low energy four dimensional theory we
have a kinetic term for this field coming from the Kahler 
potential of the form,
\begin{equation}
K_d = - \log (-i (\tau -\bar \tau))~.
\end{equation}
It is important to keep in mind that the dilaton controls the
string coupling constant, so our final minima should be
stabilized at small values of $g_s = e^{\phi}$ which in turn
means that $Im(\tau) > 1$.

One can then obtain the final expression for the Kahler
function for all the fields present in the low energy
theory as the sum of all three types of fields, namely,
\begin{equation}
K= K_{cs} + K_K + K_d ~.
\end{equation}

\subsubsection{Superpotential induced by fluxes.}

So far, we have only discussed the structure of the field
space of the four dimensional moduli, so at this level the fields
remain massless. The new ingredient recently 
added to the string theory picture is the introduction of fluxes threading the
internal cycles of the CY. These fluxes will give rise to a
four dimensional energy density that depends on the size of the
internal cycles, in other words, they will induce some potential
to the complex structure moduli \cite{GKP}. 

The expression for this
potential is obtained in the $4d$ supersymmetric theory once we know
the superpotential. This superpotential in models of flux
compactification can be shown to be of the form 
\cite{Gukov-Vafa-Witten}
\begin{equation}
W_{GVW} = \int_M{G_3 \wedge \Omega} = \int_M{(F_3 - \tau H_3) \wedge \Omega}~,
\end{equation}
where $F_3$ and $H_3$ denote the three form field strengths present
in the theory and $\Omega$ is the holomorphic three form of the CY. It
is clear from our previous definitions that this superpotential will 
depend on the dilaton as well as the complex structure moduli
therefore generating a potential for those fields.

On the other hand, the potential found this way does not depend on
the Kahler moduli so it does not help to stabilize this sector of the
theory. In order to do this one would need to include a new piece in the 
superpotential that depends on $T_i$. A model of this type was first
discussed in \cite{KKLT}.

\subsection{The KKLT construction}

In \cite{KKLT} it was argued that non-perturbative terms
would be introduced in the superpotential if there is a
gaugino condensation on a sector of the theory. The interesting
point about this new term in the superpotential is that it
would depend on the Kahler moduli therefore allowing for
the possibility of stabilizing these fields as well. A typical
term in the superpotential would have the form,
\begin{equation}
W_{NP} = \sum_i{A_i e^{-a_i T_i}}~.
\end{equation}
The KKLT construction is based on the idea of a 2 step
process. In the first part one imagines the complex structure to be 
fixed by the fluxes in a way that we described above such that their
moduli fields are stuck to their minima. The value of the 
superpotential at the minima of the complex structure fields
is denoted by $W_0$ and it is added to the non-perturbative
term coming from the gaugino condensation. Assuming
a superpotential contribution from the gaugino condensation
of the racetrack type and the presence of a single Kahler moduli for 
simplicity, the authors of \cite{KKLT} arrived
at a total superpotential of the form
\begin{equation}
W = W_0 + A e^{-a T} + B e^{-b T}~.
\end{equation}
The potential generated in this way can have several minima and
therefore fix the Kahler moduli. On the other hand, most of the minima that one would
find using this potential would have a negative value of the
cosmological constant. In particular, it is easy to see that in
supersymmetry preserving minima they would be either Minkowski \cite{KL}
or anti-de Sitter \cite{KKLT}. We must therefore {\it uplift} these minima
to be compatible with current cosmological observations.
These considerations lead them to introduce a term of the form
\begin{equation}
V_U = {D\over {{\cal V}^2}}~,
\end{equation}
where ${\cal V}$ denotes as before the volume of the CY, and
can be written in terms of the Kahler moduli $T_i$. 

Putting all these various ingredients together, one can write
the effective potential for a single Kahler moduli with Kahler
function given by $K = -3 \log \left(T + \bar{T}\right)$ in the form,
\begin{eqnarray}
\label{4d-potential}
V(X,Y) &=& {{1}\over {6X^2}} [e^{-2(a+b)X} \left(a A^2 (a X + 3)
  e^{2 b X} + b B^2 (b X + 3) e^{2 a X}\right) \nonumber \\  &+&
e^{-(a+b) X} \left(A B (2 a b X + 3 (a+b)) \cos (Y (a-b))
\right)\nonumber  \\ 
&+& 3 a Ae^{b X}  (W_{0X} \cos (a Y) - W_{0Y}\sin (a Y))\nonumber \\ 
&+& 3 b B e^{a X} (W_{0X} \cos (b Y) - W_{0Y} \sin (b Y))] \nonumber \\ 
&+& {D\over {X^2}}
\end{eqnarray}
where we have separated the real ($\text{Re} T = X$) and imaginary parts ($\text{Im} T = Y$) 
of the field as well as the superpotential ($\text{Re} W_0 = W_{0X}$)
($\text{Im} W_0 = W_{0Y}$). Using this potential one can stabilize
the Kahler moduli as well in a non-supersymmetric de Sitter vacua
compatible with our current observations \cite{KKLT}.

This is an interesting construction and there are, by now, many
different variations on how to obtain a phenomenologically
viable vacua in this way. Furthermore, as we discussed in
the introduction finding a stable compactification allows people
to think of new ways to describe the evolution of the Early 
Universe in these models. In particular, one may wonder if the
same potential that induces the compactification could be 
responsible for the energy density of inflation. Having
a computable potential away from its minimum permits us to
calculate not only the dynamics of the scalar fields in 
this potential but observational parameters such as the scale
of inflation, the amplitude of perturbations and its spectral
index \cite{Modular-Inflation}. There are many realizations of this
idea in the literature, but in the following we would like to concentrate
on a simple model of inflation, accidental inflation \cite{AI}.

\section{Accidental Inflation}

This is an inflationary scenario that is easily 
embedded in a KKLT model of flux compactification. The idea
is to look for an approximate inflection point in the Kahler field sector of
the potential that allows for sufficient inflation. In \cite{AI} 
the authors found that a region of this type could be 
generated along the real part of the complex Kahler field in the
simplest models with a racetrack type potential. 

However, one can argue that the model is quite constrained by what 
is perhaps more than one type of fine tuning. Firstly, in order 
for the inflection point to lead to a sufficient number of e-folds 
the potential has to be fined tuned so that it becomes flat enough 
around the inflection point. The hope is then that this coincidence can happen
at some point of the large parameter space available to string theory,
hence the name {\it accidental inflation}.

On the other hand, these type of models also suffer from an overshoot
problem \cite{Overshoot-Problem}. This is a
somewhat generic problem in models where the inflationary region
is small because one has to make sure that the field arrives to 
this point in space with sufficiently low velocity so it can
stay in the slow roll inflation region for sufficient amount of time. 
This problem is related to the question of what are the natural initial
conditions for the fields before inflation. This is
of course an important question for most of models of inflation 
but in these kind of inflection point scenarios this is an especially
relevant issue.

A particular example of accidental inflation can be obtained by
choosing the parameters
\begin{eqnarray}
\label{original-example}
A&=& {{1} \over {145}};~~~ B=- {{1}\over {145}};~~~ a={{2\pi} \over
  {580}};~~~ b={{2\pi} \over 600};\\ \nonumber \\ 
 W_0 &=& 1.01796 \times 10^{-4} + 3.1034287 \times 10^{-5} i;~~~ D=
 6.0614989 \times 10^{-12}~. \nonumber
\end{eqnarray}

\begin{figure}[ht] 
\centering
\includegraphics[width=100mm]{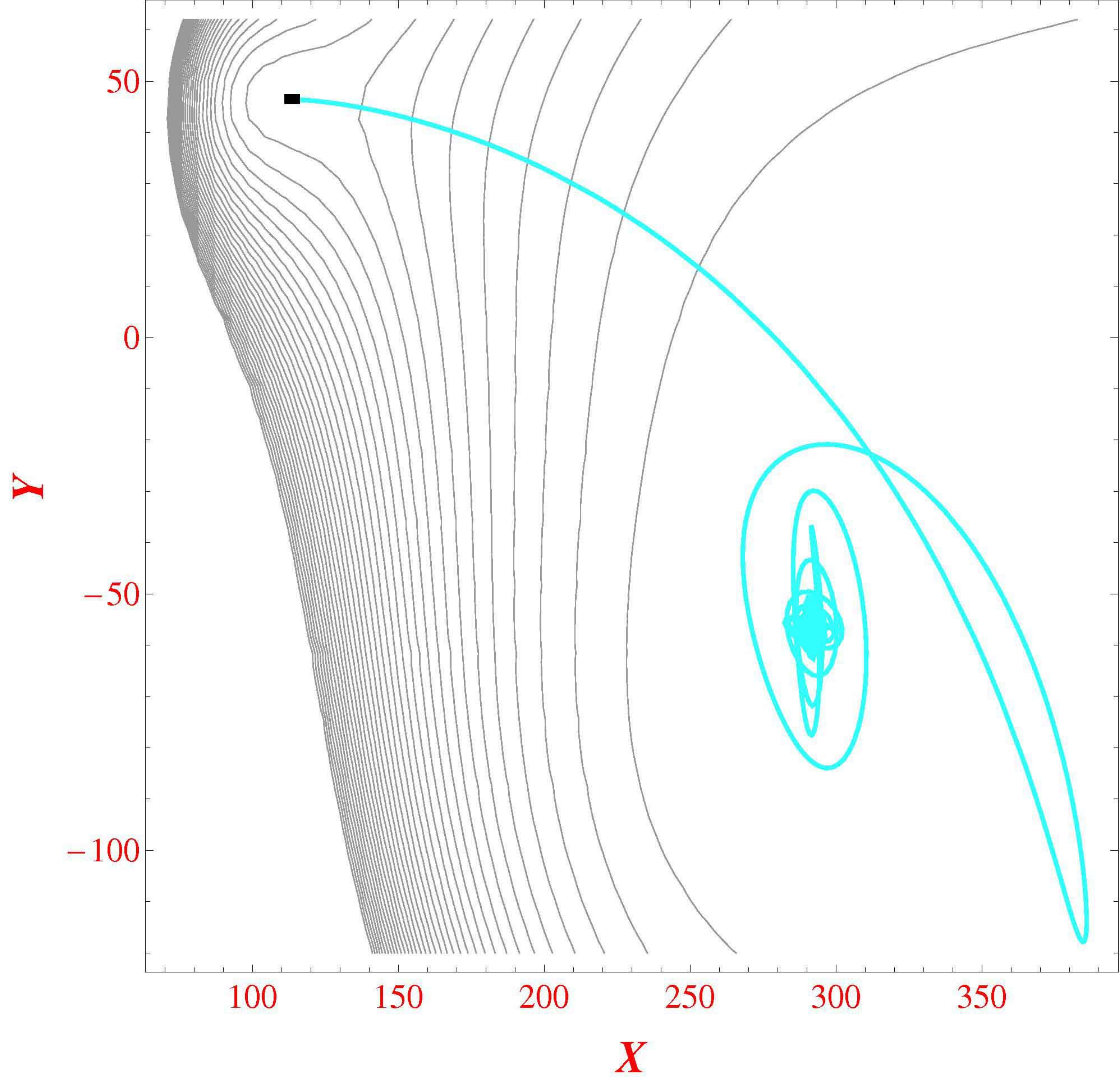}
\caption{Example of Accidental Inflation. We show the inflationary
 trajectory superimposed over the contour plot of the effective
potential in the $X-Y$ plane. We mark in black the
small region of the trajectory where the slow roll conditions are
satisfied.}
\label{FIG-1}
\end{figure}

 The idea behind this choice of values for $A,B,a$ and $b$ is that our set of parameters 
leads to an inflection point whose 3rd derivative is relatively small 
so one can have a region (not just a point) in field space that satisfies the slow roll
condition. This is not an important fine tuning since it can be
achieved in a large region of the parameter space but as we will
see this becomes quite relevant for our conclusions. We use a $W_0$ superpotential
with a real and imaginary part in order to avoid overshooting the 
overall minima of the potential, in other words to 
avoid decompactification. The effect of this complex superpotential is to displace
the value of the $Y$ component of the field at the inflection
point relative to the overall minimum. One then has a curved trajectory
in field space that allows one to reduce the kinetic energy in the
$X$ direction and avoid decompactification. We show in Fig.~(\ref{FIG-1}) the
field trajectory around the inflection point and the subsequent
evolution in a flat Friedmann-Robertson-Walker (FRW) universe where
we have chosen as our initial conditions a point in field space at the beginning of the
slow roll region. We note that even though the full trajectory is
curved, the relevant part for inflation happens near the inflection
point so the predictions of this model are closely related to 
single field inflection point models 
\cite{Inflection-point-Inflation,D-brane-inflection-1,
  Volume-modulus-inflation}.
The total number of e-folds for this case is 165 and the amplitude
of perturbations as well as the spectral index are compatible with observations.

One can easily see in this example the problems associated with the
fine tuning that we were discussing earlier. Changing the parameters in the potential
by a small amount destroys the nice properties of the inflection
point and the potential would not give rise to any number of 
e-folds. On the other hand, choosing the initial conditions
far away from the inflection point
also has an important effect since the fields pick up
too much kinetic energy by the time they arrive at the
inflection point to have enough inflation, or even worse, 
they do not approach that point
at all and run directly to the global minimum.

It is clear then that one would like to investigate the
possibility of ameliorating some of these fine tunings. In the 
following we will explore some ideas in the context of this simple
model that show how the existence of a landscape could help with 
both these issues\footnote{Note that one can also reduce 
the severity of this fine tuning problem assuming a particular family 
of measures \cite{AI} that would give a overwhelming weight to any
trajectory that inflates compared with the other ones. As described
in the introduction, we will not consider these types of measure
problems in this paper.}.

\section{Accidental Inflation in a corner of the Landscape}

As we described in the introduction, string theory
provides us with a way to scan different values of
the parameters of the low energy effective action. This
has important consequences for our understanding of what 
can be considered fine tuning of the moduli potential as well
as the predictions for the observable parameters for inflation.

 One could try to investigate this effect by assuming that the
parameters of the low energy theory would vary over some
range of values in the landscape. In our case this would mean, in
practice, to allow the first and third
derivatives of the potential around the inflection point to vary
with some prescribed distributions, similarly to what was done for 
D-brane inflation in \cite{D-brane-inflection-2,
  D-brane-inflection-3}. On the other hand
it is clear that these parameters in the moduli potential are 
not fundamental themselves but are computed in terms of the 
fundamental ingredients that vary in a quantized manner over the
landscape. In models of flux compactifications based on
Type IIB string theory, one specifies the 4d moduli potential once
the fluxes along the three cycles of the internal manifold are
fixed. This suggests that we should 
study the mapping between the various sets of fluxes and 
the values of the parameters in the low energy theory to obtain
a more accurate description of the distribution of the
parameters of the potential in a real landscape.

Given a set of the fluxes one can obtain, following the prescription 
given above, the value of the complex structure moduli 
which, in turn, fixes the value of the superpotential, $W_0$. This
means that one can think of this parameter in the potential
as being scanned over the landscape. The key point is then to 
realize that one can control the slope of the potential around 
the inflection point by choosing $W_0$ appropriately, leaving 
all the other parameters fixed. This is important for our inflationary
models since by decreasing the slope of the potential at the inflection
point one increases the number of e-folds in that region.
In the following we will use a particular CY to study in detail the distribution
of values of $W_0$ in a mini-landscape and to see how this 
affects the probability distribution of the number of 
e-fols.

There are, of course, other important fine tunings of this
modular potential that one has to address in order to obtain a successful model of
inflation in string theory. In particular one should fix the global minimum of the
potential at a vanishing value of the cosmological constant which in turn
requires us to tune the {\it uplifting} parameter $D$, so we should
also consider this parameter to be scanned over the landscape. This
is a much more serious fine tuning than the one required
for inflation to happen. We will not try to address these
two issues at the same time and in the following we will assume that 
some other sector of the landscape is responsible for the
extreme fine tuning of the parameter $D$ so we can basically
consider it a continuous parameter that can be fixed to 
have an appropriate value of the potential at its global minimum.

\subsection{The $P^4_{[1,1,1,6,9]}$ Calabi-Yau}

In order to investigate in detail the implications of the
existence of a landscape we will look at a particularly simple model for
the complex structure fields on one of the best known Calabi-Yaus,  
the orientifold of $P^4_{[1,1,1,6,9]}$ . This CY threefold has
2 Kahler moduli and 272 complex structure moduli. However
we will restrict ourselves to a 2 dimensional slice of the
complex structure moduli that can be obtained by imposing a 
particular symmetry on the manifold. (See
\cite{P11169, Better-Racetrack} for more details on this manifold). 
This is of course a very small
number of complex structure fields and it is by no means representative of a
typical CY. However we choose to work with this relatively small
number of moduli so we can explicitly perform the numerical
calculations in a reasonable amount of time. For other numerical
investigations of the complex structure sector see \cite{numerics-complex-structure}.

\subsubsection{Complex Structure}

Following the KKLT procedure described earlier, we will first find the
solutions of the supersymmetric equations for the complex structure moduli,
and the dilaton, namely:
\begin{equation}
\label{SUSY-EQ}
D_I W = 0~~~~;~~~~ D_{\tau} W = 0
\end{equation}
where $I = 1 \dots h^{2,1}$. To solve these equations we need the
information of the internal geometry, in other words, we need to find
the Kahler function for the complex structure moduli. As described 
in the previous sections the first step in this procedure is to obtain the prepotential
which in our case was computed in \cite{P11169, Better-Racetrack,modifiedP11169},
\begin{equation}
\label{prepotential}
F = (w^0)^2 {\cal F} = (w^0)^2\left({1\over 6}(9z_1^3+9z_1^2 z_2 + 3 z_1 z_2^2) - {9\over 4}
z_1^2 - {3\over 2} z_1 z_2 - {17\over 4} z_1 - {3\over 2} z_2 + \xi \right)
\end{equation}
where we use the following normalization for the complex structure fields, 
$z_i = {{w^i}\over{w^0}}$, and we have also defined
\begin{equation}
{\cal F} =  {1\over 6}(9z_1^3+9z_1^2 z_2 + 3 z_1 z_2^2) - {9\over 4}
z_1^2 - {3\over 2} z_1 z_2 - {17\over 4} z_1 - {3\over 2} z_2 + \xi~.
\end{equation}
With this prepotential and using the normalization  of $w^0 =1$, 
we can now compute the vector periods,
\begin{equation}
\Pi = (1, z_1, z_2, ~2 {\cal F} - z_1 {\cal F}_1 - z_2 {\cal F}_2,~
    {\cal F}_1,~ {\cal F}_2).
\end{equation}
Using Eqs.~(\ref{Kcs}) and (\ref{Sigma}), the Kahler function for the complex
structure moduli in terms of $z_1,z_2$ is given by,
\begin{equation}
\label{Kahler-cs}
K_{cs}(z_1,z_2)=  -\log \left[ i \left((z_1 - \bar z_1)({\cal F}_1 + \bar {\cal F}_1) + 
(z_2 - \bar z_2)({\cal F}_2 + \bar {\cal F}_2) - 2 ({\cal F} - \bar{\cal F})\right)\right]
\end{equation}
which in our case becomes, using $z_i = X_i + i Y_i$,
\begin{equation}
K_{cs}(z_1,z_2)=  - \log \left[4 Y_1 (3Y_1^2 +3 Y_1 Y_2 + Y_2^2) - 4 i \xi\right]
\end{equation}
following \cite{Better-Racetrack} we use $\xi = - 1.3 i$.

The superpotential generated by the fluxes can be computed using the expression
\begin{equation}
W_F = \frac{1}{(2\pi)^2\alpha'} \int_M{(F_3 - \tau H_3) \wedge \Omega} 
\end{equation}
which can also be written in terms of the dilaton and complex
structure as well as the integer fluxes through the $A$ and $B$ 3-cycles as,
\begin{equation}
\label{superpotential-cs}
W_F= \sum_{i=0}^2{\left[(f^i_{A} - \tau h^i_{A}) F_i - (f^i_{B} - \tau h^i_{B}) z_i\right]}
\end{equation}
where we have defined
\begin{equation}
f^i_{A,B} = \frac{1}{(2\pi)^2\alpha'} \int_{A^i, B^i}{F_3} \in Z~,
\end{equation}
\begin{equation}
h^i_{A,B} = \frac{1}{(2\pi)^2\alpha'}  \int_{A^i, B^i}{H_3} \in Z~.
\end{equation}
Fixing the flux integers and using the definitions of the Kahler function,  
Eq.~(\ref{Kahler-cs}), and the superpotential,
Eq.~(\ref{superpotential-cs}) we can now obtain the 
values of the complex structure fields and the dilaton at a
supersymmetric minima by solving Eqs.~(\ref{SUSY-EQ}). Plugging
the solutions back into the expression for the superpotential we can 
easily compute $W_0$ at that point.

\begin{figure}[ht] 
\centering
\includegraphics[width=120mm]{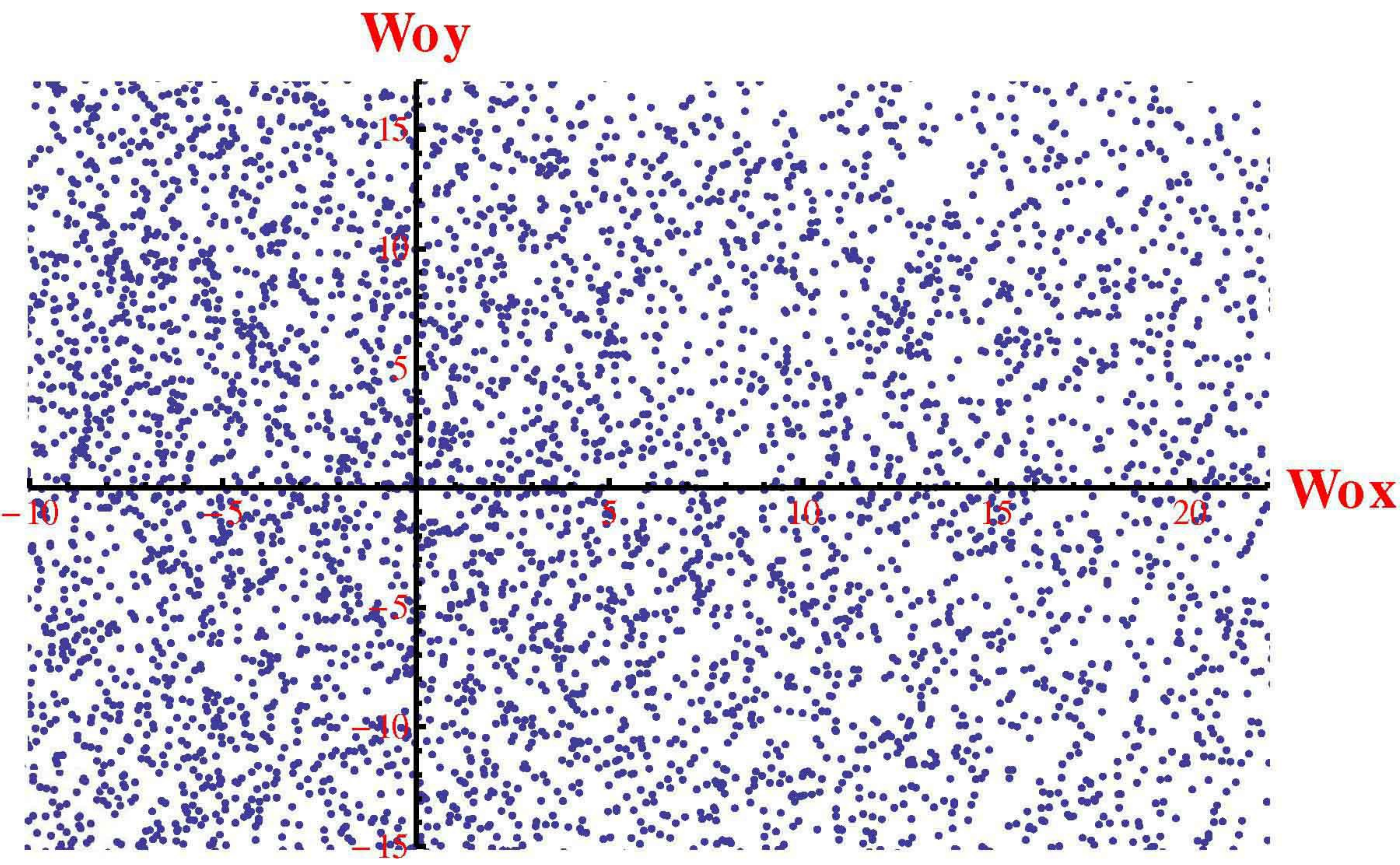}
\caption {Values of $W_0$ for a small sample of supersymmetric
  solutions in the landscape generated by turning on six fluxes over
  the $P^4_{[1,1,1,6,9]}$ manifold. }
\label{Uniform-W0}
\end{figure}

We have followed this procedure for a large number ($\approx 10^9$)
of combinations of the fluxes and observe that the values of
$W_0$ seem to be uniformly distributed over a large area (of
the order of $10^4$) of the complex plane. 
We show in Fig.~(\ref{Uniform-W0}) a small sample of $10^4$ randomly
selected values over a small region that clearly demonstrates this point.
This is a similar result to the one obtained in \cite{Douglas} and
it will have important consequences for us later. 
Note that not all these points will end up being minima of the full
potential, since some of them may be stable supersymmetric
AdS critical points which will be transformed into saddle points
by the uplifting procedure. One can, of course, resolve this issue
by identifying the true minima of the full potential, but for the
moment we will be content with this KKLT procedure. Our results
should not change even if the fraction of these points
that are true minima is small.

\subsubsection{The Kahler moduli}

We are mainly interested in the effect of the complex structure
moduli in our inflationary model so in order to simplify our
analysis we will focus on a very simple toy model for the 
Kahler fields whose Kahler function is given by
Eq.~(\ref{Kahler-moduli-Kahler}). 
We will also assume the presence of a non-perturbative potential as well
as an uplifting term as described in the KKLT constructions
above.\footnote{Note that one could
take the Kahler moduli specific for our CY manifold, but this 
would make the inflationary model considerably more complicated. We
will come back to this issue as part of our future work.}.

Finally we will also consider the case where the parameters of the non-perturbative 
superpotential $A,B,a,b$ do not have a strong dependence on the 
complex structure moduli and take them to be constant for all the
values of the fluxes. This is likely to be true for many of our
vacua since we have not seen large changes of the values
of the complex structure moduli over the scanned vacua. Even if this
assumption is violated in some cases, it will be true for many
of the vacua and our conclusions are likely to hold.

\subsection{Exploring a mini-Landscape}

In the previous section we have shown an example of a successful 
inflationary model where all the observational constraints 
of the model were satisfied. In particular one could find a region
of the potential that allowed for a large number of e-folds. This was achieved by a 
quite severe fine tuning of the superpotential around  
$W_0 =  1.01796 \times 10^{-4} + 3.1034287 \times 10^{-5} i$.
This implies that the possible values of $W_0$ to make this
model work would be confined to a tiny area in the $W_0$ complex plane of the order of
$10^{-17}$. Taking into account that the distribution
of vacua on the $W_0$ plane seems to be uniform, we can estimate 
the fraction of vacua that would be found in this preferred region 
if one was to generate a large number of flux combinations. 
Following this calculation we can easily see that we will not 
be able to explore the landscape finely enough
with our mini-landscape in a reasonable amount of time.

On the other hand, the main reason to go to these small values of
$W_0$ was to satisfy all the observational constraints, in particular 
the idea that the scale of inflation should be small to accommodate the
amplitude of perturbations. In the 
following we will sacrifice this requirement in order to demonstrate 
in a concrete example some of the ideas presented earlier about the
fine tuning necessary to achieve large number of e-folds. We therefore 
explore another region of the parameters where one still needs a 
considerable degree of fine tuning but can be accessed with our 
mini-landscape with a sample of generated
vacua of a much more manageable size.

\subsubsection{Specific Example}

We start by picking a generic set of values for $A,B, a$ and $b$ in a 
region of parameter space that
leads to a large value of $W_0$. We imagine that these parameters are
fixed by the effective theory of a hidden sector. We take
the values,
\begin{equation}
\label{EXAMPLE}
A= 5; ~~~ B=- 10; ~~~ a={{2\pi} \over {100}}; ~~~ b={{2\pi}
  \over 290}~.
\end{equation}
One can now obtain a range of values for $W_0$ and $D$ such that
the potential would have a near inflection point at some positive
value of the potential, as well as a global minimum with vanishing
cosmological constant. This requires $W_0$ to be within a small area 
in the complex plane of the order of $\approx 10^{-4}$. This is of
course a fine tuned value from the low energy perspective,
but the question we would like to address is if one should expect
to find several vacua within this narrow strip of values or not,
taking into account the existence of the landscape.

\begin{figure}[ht] 
\centering
\includegraphics[width=120mm]{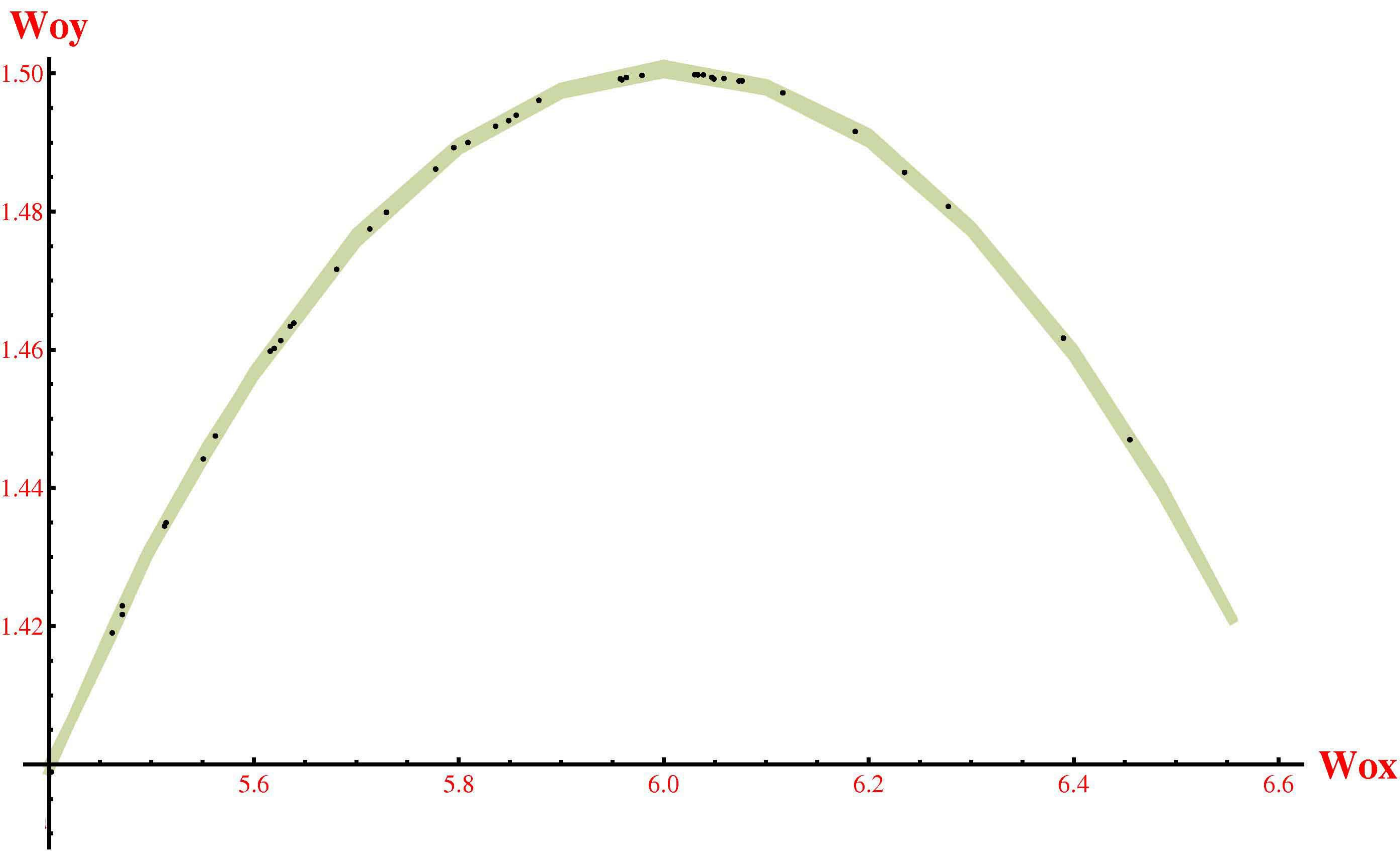}
\caption {We show in this figure the small shaded area in the $W_0$
  plane consistent with more than $60$ e-folds of inflation for the
  parameters given in Eq.~(\ref{EXAMPLE}). Each of the points in this region
represent a particular combination of fluxes that leads to
this value of $W_0$.}
\label{inflection-strip}
\end{figure}

In order to address this question we find the values for the
superpotential at the supersymmetric minima using the machinery
described above. We do this by first generating 
a large number (of the order of ~$10^9$) of combinations of fluxes
and solving the supersymmetric conditions Eqs.~(\ref{SUSY-EQ}). Armed
with the values of the complex structure and dilaton at their minima,
we calculate the superpotential at those values
and identify the ones that land within our region of
interest. Following this procedure, we were able to find $\sim 50$
combinations of fluxes with the correct values of the superpotential.
We show in Fig. (\ref{inflection-strip}) the region of $W_0$ required to obtain 
inflation in this model as well as the location of the particular
values of the flux vacua the we found following the steps described 
above.

We note that the number of vacua we found in this region is actually 
in good agreement with the assumption that the
superpotential scans uniformly the values of the complex plane
within the area between roughly $W_X =(-100,100)$ and $W_Y = (-100,100)$.
This suggests that one can follow this calculation for other
cases that can not be explicitly done by numerical calculation
and therefore one can assume that it is likely that the small
fine-tuned region of the $W_0$ is actually achieved by quite
a large set of flux vacua.

The values of the fluxes are not totally arbitrary, they are
constrained by several technical requirements. The first one is 
due to the tadpole condition, which basically imposes
that the total $D3$ brane charge be zero. In practice,
this creates a upper limit on a combination of the fluxes
over the internal manifold. Here we will follow Douglas et
al. \cite{Better-Racetrack} and will only consider the sets of 
fluxes restricted to the condition,
\begin{equation}
N_{flux} = {1\over\left( 2 \pi \right)^4  \alpha^{'2} }  \int_M{F_3
  \wedge H_3} < 350~.
\end{equation}

Also, we will only keep fluxes whose vacua are stabilized at a
large value $\text{Im}(\tau)$. This will make sure that we will find these
vacua in the weak coupling regime.
Similarly, in order to disregard instanton corrections in the
prepotential in Eq.~(\ref{prepotential}) we will only use fluxes
that lead to $\text{Im}(z_i)\geq 1$ for $i=1,2$. Finally we are only 
considering fluxes that are inequivalent, meaning
our set of the fluxes are not related by a $SL(2,Z)$ symmetry 
transformation. This prevents us from overcounting the number
of fluxes in the region of interest for us.\footnote{ These
  requirements 
reduce the number of valid minima from the 
naive calculation based on the uniform distribution of vacua
on the $W_0$ plane since, sometimes, many of these vacua violate
some of these restrictions. We have not seen however that
this would produce any voids in the $W_0$ distribution, so the
argument is still basically valid.}

\subsubsection{Initial Conditions}

In order to make predictions for the observable 
parameters of this landscape one has to consider some initial conditions.
As we argued above choosing these type of values of the parameters
relaxes the extreme fine tuning of the
initial conditions in this model, since we now have a flat section of
the potential where the slow roll conditions can be satisfied, 
but this does not explain why
should the universe start at all close to the inflationary
plateau in the vast region of field space.

 Here we point out that the idea of the landscape also helps us understanding why
this happens. Let us think for a moment on the other flux 
vacua in the theory, the ones that could be the parent vacua for
the one that we find ourselves today. It is clear that there will be many
other combinations of fluxes that give a nearby value of
$W_0$. The important point is to realize that many of those other
values of the superpotential will turn the
inflection point in the Kahler moduli potential into a local minima. 
This makes it possible for the fields to be stuck on a particularly
interesting value in the parent vacua, somewhat near the inflationary plateau of the
daughter vacuum \footnote{Similar ideas were also studied in the
  context of D-brane inflation in \cite{Dynamical-fine-tuning}.}. 
One can imagine that the flux changing transition
would mainly affect the complex structure fields and would not 
have a great impact on the values of the Kahler moduli. This is a
reasonable assumption since, after all, this
transition could happen in a local part of the internal geometry
and presumably would not change the overall volume of the 
internal manifold (parametrized by the field $X$) by a whole lot.

Having this process in mind,
one should consider that each of these satisfactory {\it daughter} vacua
can have in principle many predecessors that can give rise to it,
({\it  her parent vacua}). The
idea is then that the initial conditions for the field evolution
in the daughter vacuum should be set by the conditions in the 
predecessor. This suggests that we should look for the form of the
effective potential outside of the region of the $W_0$ that gives
an inflection point inflation and identify the minima of that 
other vacua. In order to do this in our example, we choose one 
particular daughter vacuum and investigate it in more detail, assuming 
that we only change the value of $W_0$, in
other words we will leave the parameter $D$ constant. 

For example, let us consider the following set of flux integers,
\begin{eqnarray}
f^i_A &=& (17,-2,0)~~~;~~~ f^i_B = (5,-47,-12)~; \\
h^i_A &=& (-2,-4,4)~~~;~~~h^i_B = (44,22,3)~.\nonumber
\end{eqnarray}
With these fluxes one can show that the solution of the supersymmetric
Eqs.~(\ref{SUSY-EQ}) for the complex structure 
moduli and the dilaton take the values,
\begin{eqnarray}
z_1 &=& -0.749 + 0.991 i ~, \\
z_2 &=& 2.043 + 0.977 i ~,\\
\tau &=& -1.28 + 2.87 i~.
\end{eqnarray}
Using these results we obtain the superpotential at this point,
\begin{equation}
W_0 = 5.87805764 + 1.49611588  i~~,
\end{equation}
while the uplifting parameter in this case should be,
\begin{equation}
D=0.0642811355~.
\end{equation}
Taking all these values into account we use the scalar potential in
Eq.~(\ref{4d-potential}) to find the inflection point in the Kahler
moduli fields at,
\begin{equation}
 T=7.8623431 +  14.2923151 i ~,
\end{equation}
as well as the global minimum which in this case is situated at
\begin{equation}
 T=66.785459 - 15.343517 i~.
\end{equation}
One can show that this potential qualifies as a successful
daughter vacuum leading to roughly $115$ e-folds of inflation.
Changing slightly the value of the superpotential from 
this one would make the inflection point region steeper or transform it
into a local minimum. Thinking in terms of the complex $W_0$ one can
see that there is a relatively large area where one would
find a de Sitter local minimum.\footnote{We will not consider 
the AdS minima as possible parent vacua since they would 
likely be collapsing before they have time to tunnel to other
flux vacua.}

\begin{figure}[ht] 
\centering
\includegraphics[width=120mm]{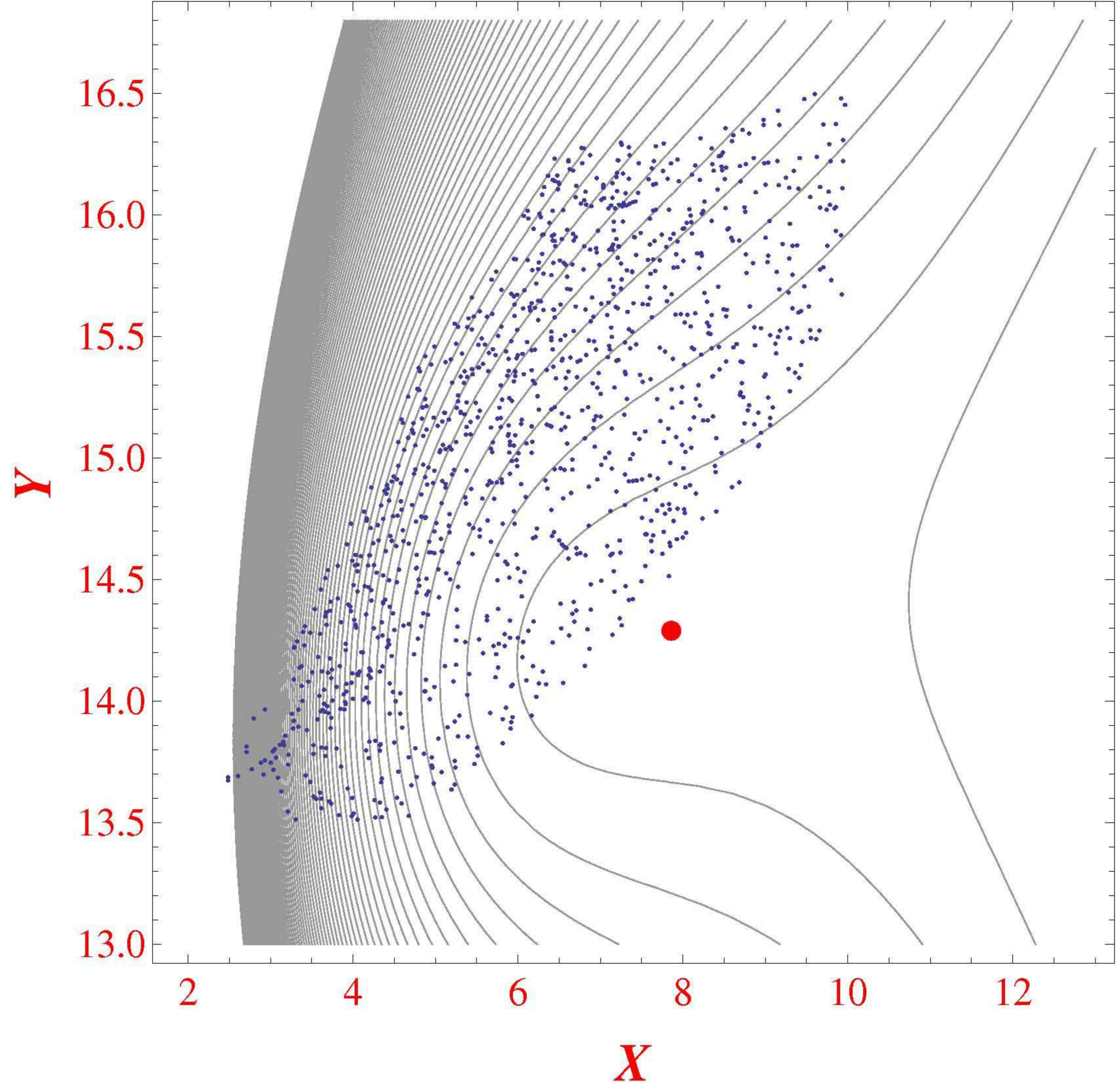}
\caption {Plot of the location of the de Sitter parent minima around
  the inflection point (big red circle) of the daughter
  vacuum in the $X-Y$ field space. We show in the background the
contour plot of the potential for the daughter vacuum case. }
\label{dS-vacua}
\end{figure}

On the other hand each set of fluxes gives, at the supersymmetric minimum, a value for 
the superpotential with a different complex phase and therefore it
shifts the inflection point or the minimum in the $X-Y$ field
space. We show in Fig.~(\ref{dS-vacua}) the contour plot of the potential for the set
of parameters of our {\it daughter} vacuum together with the
location of a few hundred de Sitter minima that we found
by varying the combinations of the fluxes. We consider any
of these points a good location for the initial conditions for
the interior of the {\it daughter} bubble that forms as a result
of the quantum tunneling event. We note that 
these are not nearby vacua in the sense of a normal metric
on the space of fluxes. In fact, some of these vacua may be
away from our daughter vacua by changes in several fluxes. It would be
interesting to study the distribution of decay rates for this
set of vacua along the lines 
of (\cite{KPV,FLW,BPSPV, Tunneling-with-monodromies})\footnote{One
  should also consider other possible ways to induce multiple flux
  transitions that may be relevant here. See for example
  \cite{Multiple-steps, Flux-discharge}.}. 
This is important since it enters the final calculation of the
probability distribution of any observable in the multiverse
\cite{Guth-Nomura}. We leave this important issue for future work.

This picture suggests that many of the predecessor
vacua for this model would be situated near the inflection
point for our daughter vacua. This does not solve all our problems, 
since even if we start our cosmological
evolution from those points, we will still have to
face the overshooting problem. Here we argue that 
the idea that the inflationary regime in our past was initiated by a
flux changing transition also helps with this problem. 

It was pointed out in \cite{FKRMS,DVW,VW} that the presence of a curvature
dominated regime in the early stages of the
interior of a newly created bubble could help solving the 
overshooting problem by gently depositing the fields over
the inflationary plateau. The situation is 
 more complicated in our case, since we have these parent
vacua scattered over the $X-Y$ plane which seems to make the problem
of dynamically finding the inflection point a little bit harder.

\begin{figure}[ht] 
\centering
\includegraphics[width=120mm]{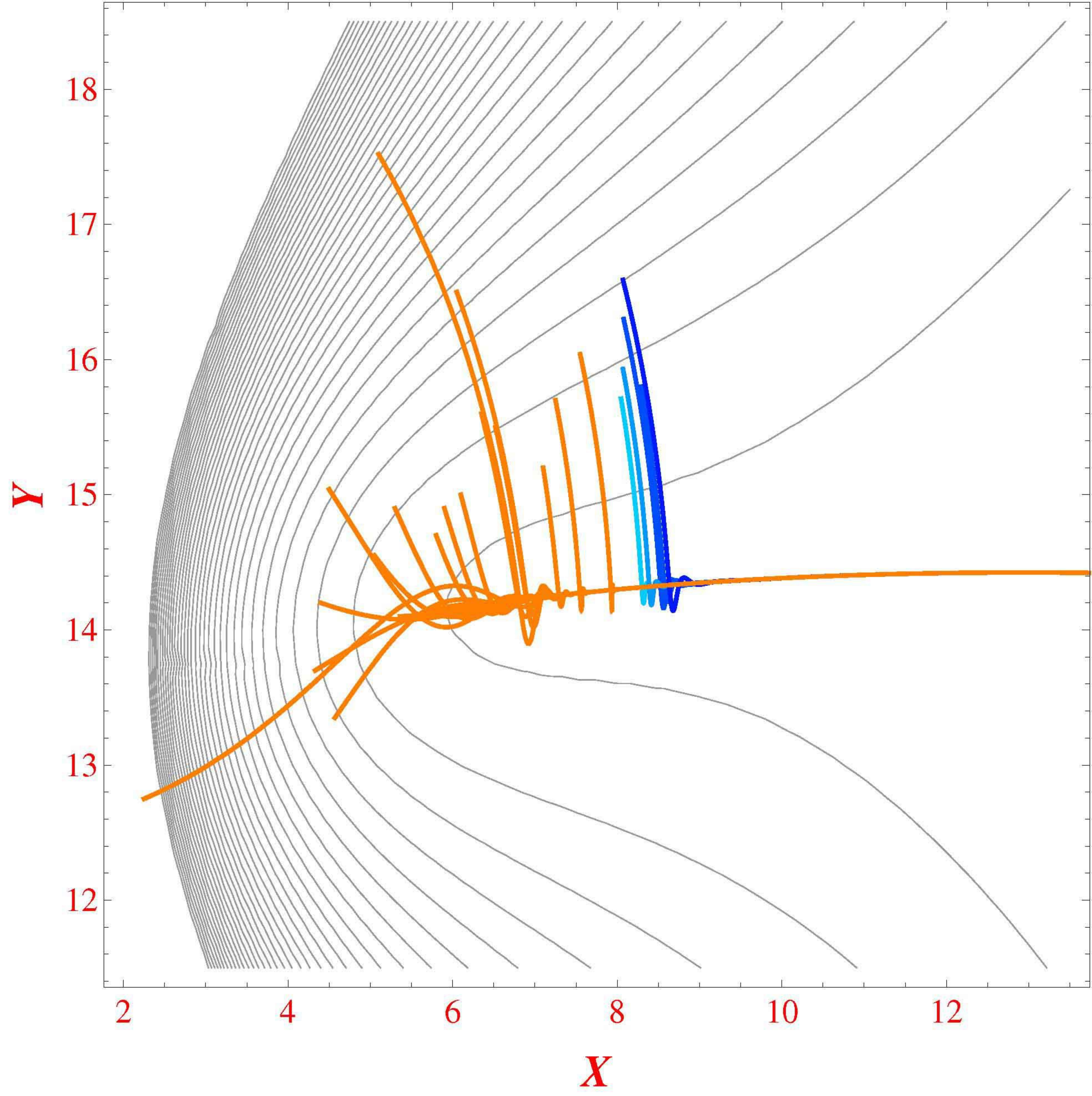}
\caption {Plot of a small number of inflationary trajectories for
  different initial conditions around 
the inflection point of the daughter vacuum. All the trajectories
converge to the same path at the inflection point demonstrating the
attractor-like behaviour in our model.}
\label{trajectories}
\end{figure}

In order to investigate these ideas we take a large number of de Sitter parent vacua for our
case and find, using the equations of motion given in the Appendix, the evolution of
the Kahler moduli in the open FRW universe inside of the
bubble. We see that even though the initial point is in some cases
far away from the inflection point, the fields roll towards it without 
overshooting it. This is due to a combination
of effects. The first one is the one that we described earlier, the
help of friction coming from the fact that the universe is open. 
The second effect is the evolution of the fields along the
perpendicular direction, $Y$. This evolution allows for some dissipation
of the energy stored in the initial conditions and helps the fields
to arrive at the slow roll region without so much kinetic energy. The result
is an attractor-like behaviour towards the inflection point that is
easily seen in Fig~(\ref{trajectories}). This is an important effect since it 
will increase the range of possible initial conditions that one could 
take in order to have certain number of e-folds.

\subsubsection{Distribution of the number of e-folds}

One of the interesting questions we can address in this mini-landscape
is what is the distribution of the number of e-folds 
within the inflationary {\it daughter} vacua. This was
studied for a simple model of the landscape in \cite{FKRMS} where
it was calculated to be a $1/N^4$ distribution. We would like to
understand the similar situation in our case taking
into account the inflationary daughter vacua found in 
Fig.~(\ref{inflection-strip}). To calculate this
distribution, we want to consider the two effects present here, the fact that
the effective potential changes with the value of $W_0$ as well as the
possible effect of the distribution of the initial conditions.

\begin{figure}[ht] 
\centering
\includegraphics[width=120mm]{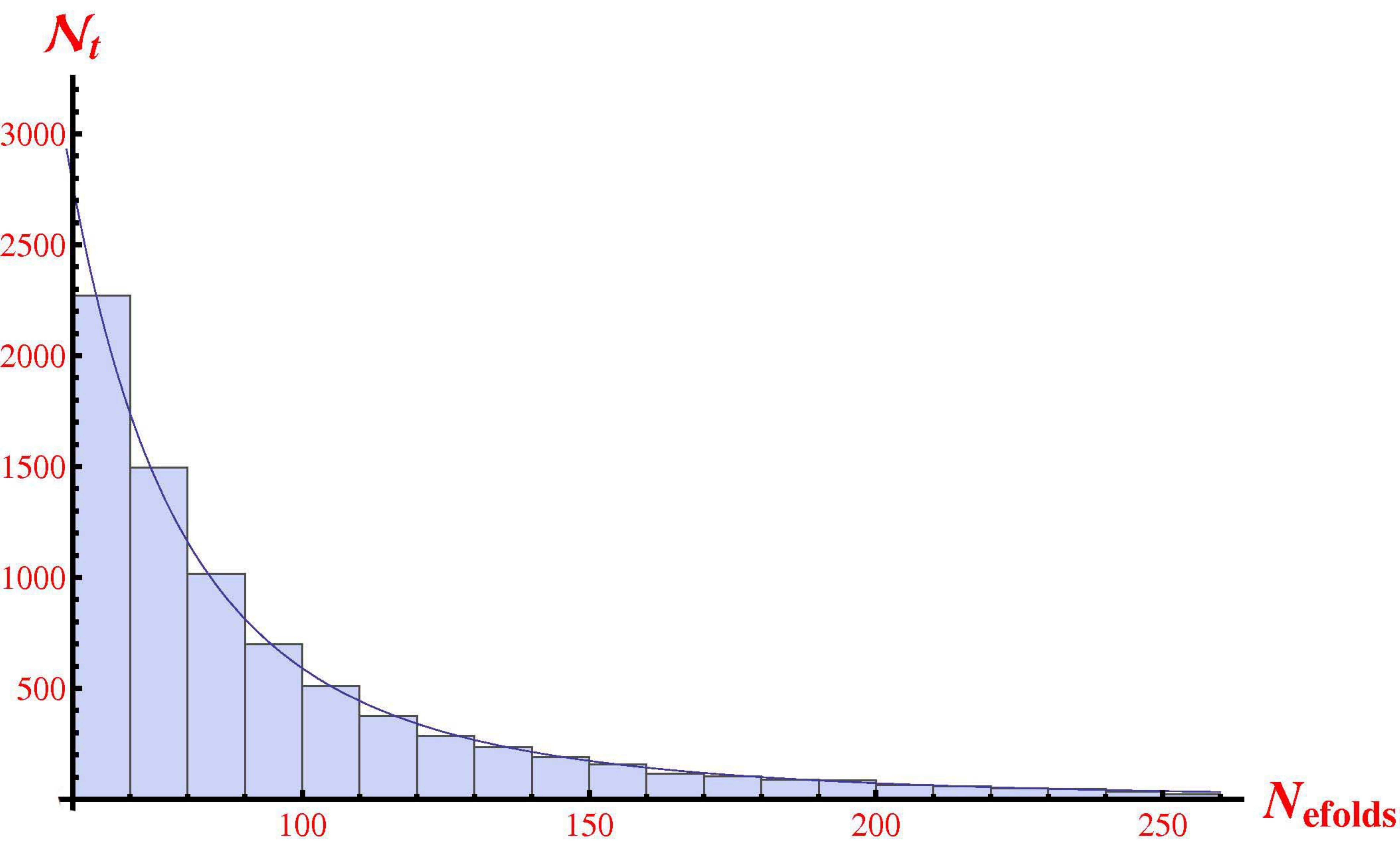}
\caption {Distribution of the number of e-folds. We show the histogram of the number of trajectories
  (${\cal N}_t$) as a function of the number of e-folds, ($N_{\text{efolds}}$).}
\label{NE-in-the-landscape}
\end{figure}

We investigate this by looking at the evolution of the fields in each 
of the realizations of $W_0$ by using some random initial conditions  
near the inflection point \footnote{We could, in principle, use
the exact location of the de Sitter vacua by calculating the
position in each case, but we simplify things a little bit here by
taking random initial conditions since the distribution in field 
space is pretty homogeneous.} as well as the assumption of an open 
universe. We show in Fig.~(\ref{NE-in-the-landscape}) the histogram of 
the number of e-folds for this set of vacua. 

The results are well approximated by a $1/N^3$ distribution. One can
explain this behavior observing that the distribution of the first derivatives at the 
inflection point is flat in this ensemble of vacua and assuming that,
due to the attractor like behaviour, the
initial conditions do not play a significant role in this
distribution. This is a similar result to the one obtained in
\cite{D-brane-inflection-2, D-brane-inflection-3} although in our case we have obtained this
distribution directly from the fluxes, and it is not
an assumption about the distribution of values of low energy
parameters. We give the details of this derivation in a one
dimensional toy model in the Appendix.

If this was the only observable prediction of this
landscape we would be tempted to argue that it is quite
likely to see a small amount of curvature in the universe
today, since large numbers of e-folds are hard to
achieve inside of our bubble universe. This was first discussed 
in \cite{FKRMS} in a simple toy model for the landscape. (See also the discussion in
\cite{Salem}). 

The situation is more complicated in
our case if we consider the constraints obtained from the
cosmological perturbations associated with the inflaton field.
However, we will not discuss the distribution of values of these other observables with the
present set of parameters since we are considering a corner of the
landscape where the scale of the potential is too
high. Remember that this was the prize we had to pay in order to
investigate actual vacua of the complex structure minima directly from
the fluxes. In the following section, we will return to our original example 
where we do not have this problem.

\section{Other observable parameters in the Landscape}

 We can now extrapolate the results of the previous section to other
regions of the landscape that we can not directly access numerically
since the number of required vacua that we would need to explore
would be enormous. In particular, we can investigate the dependence of other
observational parameters like the amplitude of perturbations as well 
as the spectral index in the phenomenologically viable model given 
by Eq.~(\ref{original-example}). In order to proceed we
will assume that the distribution of values of $W_0$  is uniform
over the landscape and dense enough in our region of interest and
that there are many minima nearby in field space to our inflationary
inflection point.

We numerically evolve a large number of inflationary
trajectories assuming a random initial condition for the fields near
the inflection point in a potential generated by choosing a random value
for $W_0$ within the tiny area compatible with more than $60$ e-folds.

\begin{figure}[ht] 
\centering
\includegraphics[width=140mm]{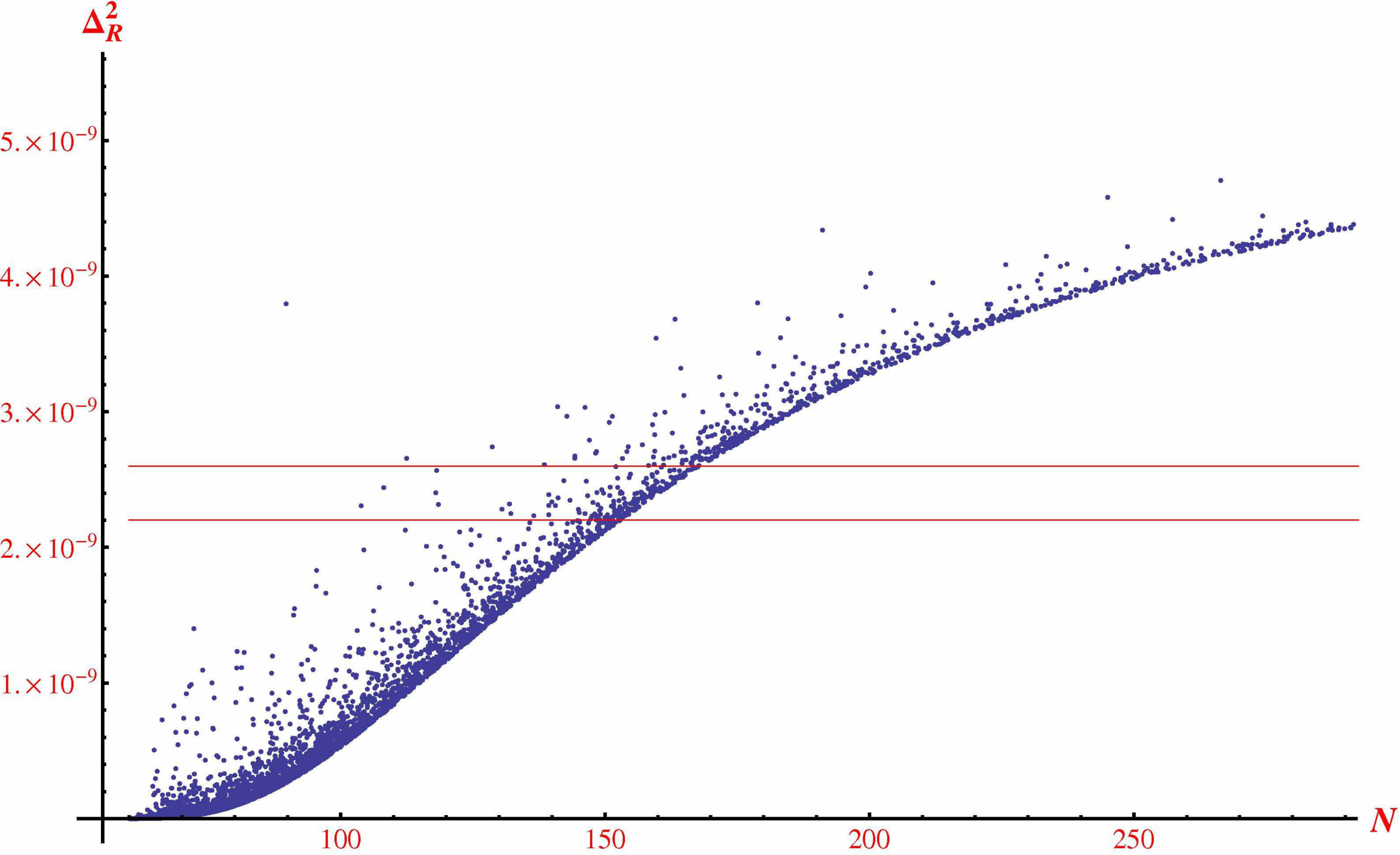}
\caption {Values for the amplitude of perturbations $60$ e-folds
  before the end of inflation as a function of the total number of
  e-folds for the simulated trajectories in our mini-landscape of 
accidental inflation.}
\label{Amplitude-versus-N}
\end{figure}

\begin{figure}[ht] 
\centering
\includegraphics[width=140mm]{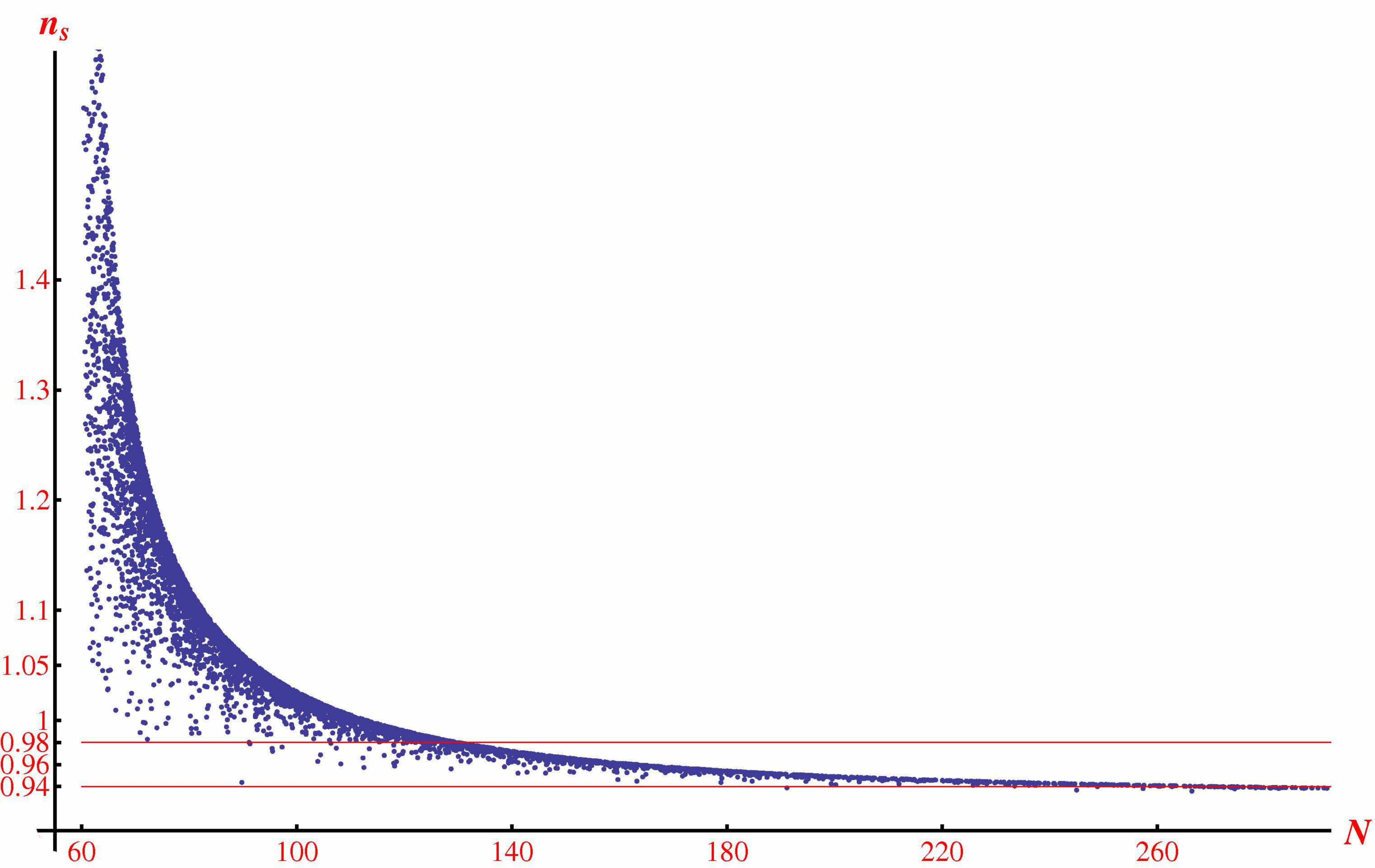}
\caption {Values of the spectral index ($n_s$) versus the number of
  e-folds for our mini-landscape.}
\label{ns-versus-N}
\end{figure}

We plot in Fig.~(\ref{Amplitude-versus-N}) the
amplitude of scalar perturbations found $60$ e-folds before the end of
inflation, on a run of $6000$ different
realizations together with the narrow
band of the $2\sigma$ deviation from the observed value \cite{WMAP7}. 
We see that this imposes a pretty strong constraint on the possible trajectories and allows us to 
discard many of them. We then proceed to calculate the
spectral index predicted in this case and we show our 
results in Fig.~(\ref{ns-versus-N}) as well as the
$2\sigma$ experimental band observed by WMAP.

We see on these two figures what seems to be a 
strong dependence of the observables on the 
number of e-folds together with some scattered
points around it. This is again a manifestation
of the fact that the most important effect that one
introduces by changing the $W_0$ is 
to modify the slope of the near-inflection point. 
Most of the trajectories for each individual
potential are close to the attractor solutions given by
the single field slow roll conditions. Assuming these two effects
 one can account for the general dependence of these observables with 
the number of e-folds. We show how this occurs in the Appendix
for a single field toy model.

Finally, the distribution on the number of e-folds in this
case is again well described by a $1/N^3$ dependence reinforcing
the idea that we can think of this landscape as being dominated
by a flat scanning of the first derivative of the inflection
point inflationary potential.

We conclude that only $4\%$ of all our trajectories
are compatible with the current observational
constraints. The main reason for this is that most
of the trajectories have a small number of e-folds
and a blue spectrum, as one would expect for an
inflection point \cite{Inflection-point-Inflation,D-brane-inflection-1,
  Volume-modulus-inflation}. The results for those viable cases are highly peaked
around the attractor solution with $N_{efolds} = 160$, $n_s = 0.96$
and $\Delta_{\cal R} = 2.5 \times 10^{-9}$, but there are a very
small number of trajectories that correspond to the edges of the
basin of attraction of this solution. An example of this
would be a trajectory that started far away from the
inflection point and reached the $60$ e-folds before the end
of inflation mark at the end of the slow roll region having undergone a small number of 
e-folds. These are interesting solutions where one may be
able to observe some curvature. On the other hand, they
are highly subdominant.

One could of course imagine a curvaton
type scenario where the cosmological perturbations are generated by a 
different field not related to the inflaton. This is certainly
a possibility that one could study in a string
theory setup, see for example \cite{Cicoli}. Following these
ideas one decouples the distribution of the number of e-folds
from the other observables related to the perturbations
which can have an important effect on the overall predictions on
the observable parameters in this landscape.

It is also important to emphasize that these results are 
obtained assuming the same values of the parameters
$A, B, a$ and $b$. In practice, this means that we
are exploring a particular sub-sector of the
landscape with a fixed hidden field theory. 
One can imagine that these parameters could 
also be scanned over in different
sectors of the landscape. Changing the
scale of, for example, $A$ and $B$, would directly 
affect the scale of inflation so in principle one
can rescale the amplitude of the perturbations to 
include some trajectories and not others.

\begin{figure}[htbp] 
\centering
\mbox{\hspace{-.2in}
\subfigure{\includegraphics[width=3.1in]{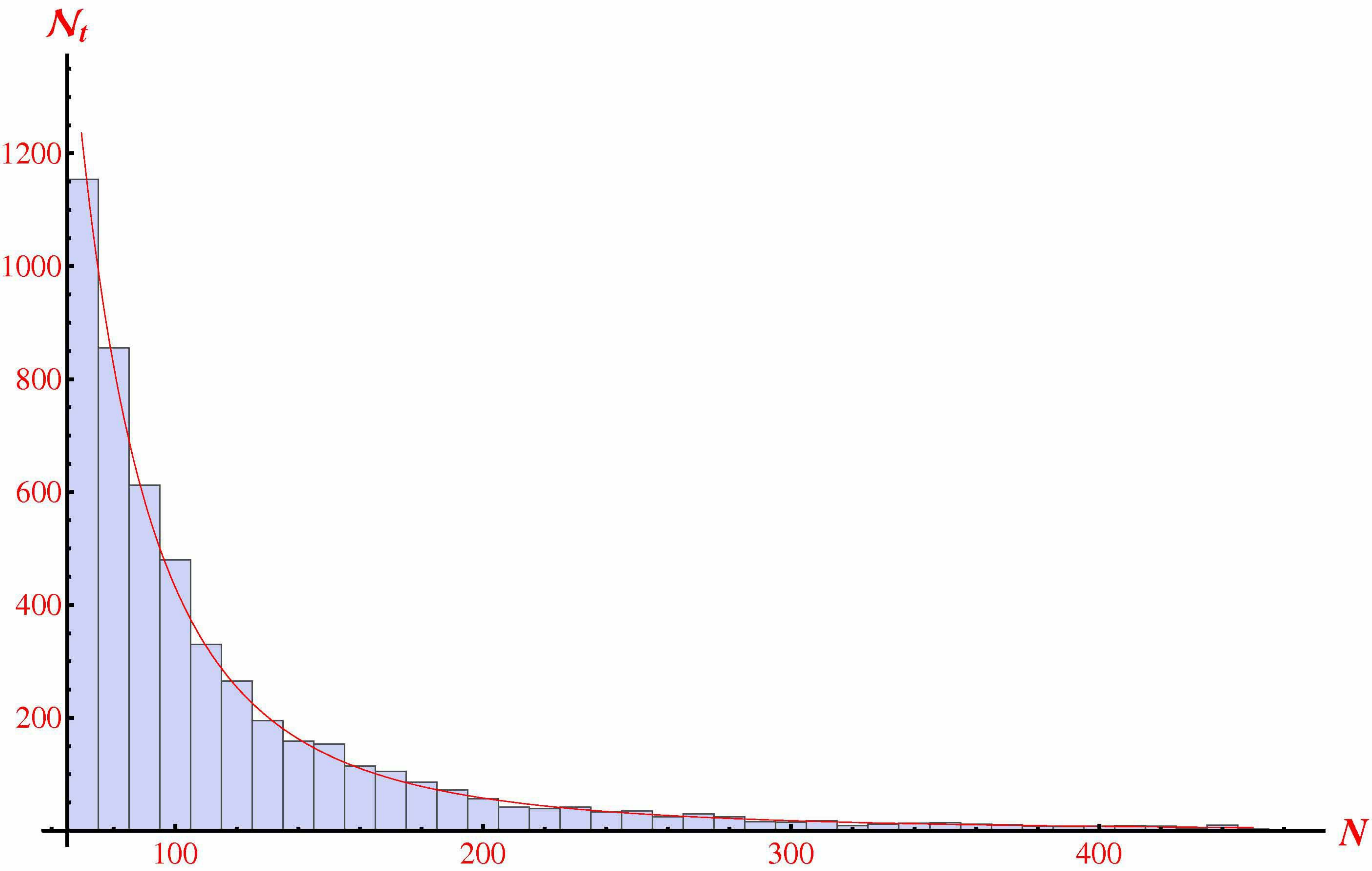}} 
\subfigure{\includegraphics[width=3.1in]{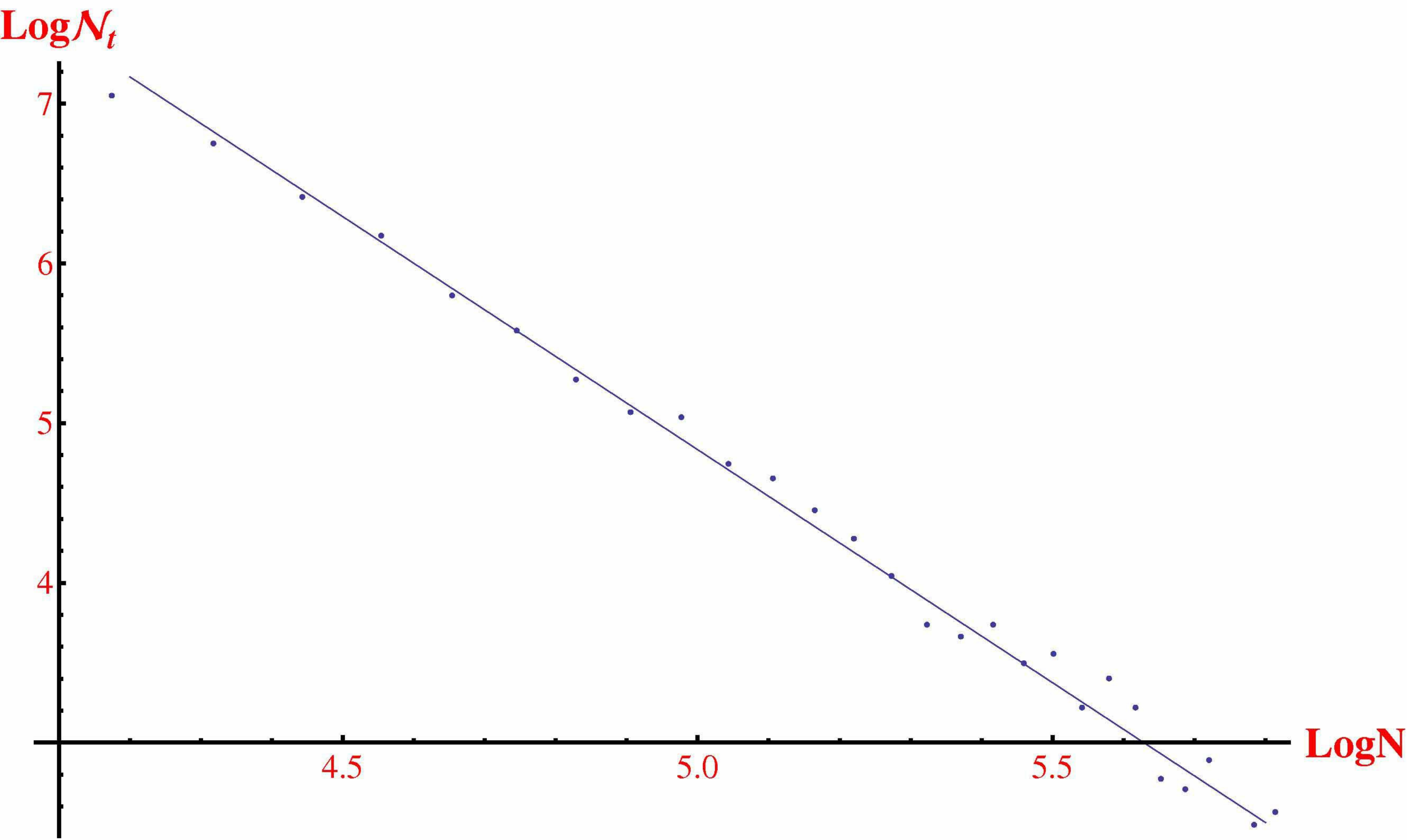}}}
\caption{Distribution of the number of e-folds in our simulated
  landscape. We show on the right hand figure the best fit of the data
to a curve of the form $P(N) \sim N^{\alpha}$ with $\alpha = 2.92 \pm 0.06$..}
\label{PofN}
\end{figure}

\section{Conclusions}

We focused in this paper on a particular model of inflation named
Accidental Inflation \cite{AI} where the potential has to be 
fined tuned in order to give a substantial number of
e-folds. We showed that this apparent fine tuning 
can be generically obtained by scanning the
form of the potentials found in a very modestly small
sector of the landscape generated by a family of six fluxes.
Furthermore, the existence of a landscape in this model
provides us with a theory of initial conditions for the
inflationary period. Changing the fluxes from the 
cosmologically interesting one (the one that we have
recently followed in our past history) one sees that the potential
develops a local minimum nearby in field space. This is
also a generic situation and we can show that there are
many other vacua of this kind nearby. This suggests the scenario where
the universe evolved from one of these vacua by
tunneling out of it by a flux-changing instanton that
triggers the transition to the daughter vacuum. This process 
gives us a natural way to select good initial conditions for
our subsequent evolution avoiding overshooting problems
that normally occur in these type of models. 

We have made some approximations in this
paper and it would be interesting to investigate the
extent of their validity. In particular, one can 
improve the calculation presented here by incorporating
the actual Kahler moduli fields for the $P^4_{[1,1,1,6,9]}$
manifold. This would also allow us to explore
other kind of models like the Large Volume Scenario \cite{LVS}
and possibly find minima of the complete supergravity potential
and not rely on the KKLT type of constructions. Another
interesting point would be to incorporate the dependence
of the parameters $A,B$... on the complex structure moduli.
This is important since it would likely affect the conclusions
about the distributions of the scale on inflation in a
particular model.

We have also neglected other possible corrections
to the potential that could be important for inflation.
One can try to repeat the calculation that we have
performed here taking into account those
other terms estimated in \cite{Corrections}. 

Finally, one would like to extend this kind of 
arguments to other models of the string cosmology
in order to be able to draw more generic conclusions
that in combination with a measure would lead to
a prediction of the inflationary observables in the
string landscape.

\section{Acknowledgements}
It is a pleasure to especially thank Per Berglund for his help with
several important technical points. We would also like to thank 
Ken Olum, Delia Schwartz-Perlov, Mike Salem, Alex
Vilenkin and Ben Shlaer for many useful discussions. J.J.B.-P. 
is supported in part by the National Science Foundation 
under the grant, PHY-0969910. M. G-R is supported by the 
Ram\'on y Cajal programme through the grant RYC-2009-04433. K.M. 
was supported in part by the John F. Burlingame Graduate Fellowship
at Tufts.
\appendix

\section{Equations of motion for $N=1$ supergravity}

Starting with the ${\cal N}=1$ supergravity action
\begin{equation}
S=-\int d^{4}x\sqrt{-g}\left[\frac{1}{2}R + K_{I \bar J} \partial_{\mu} \Phi^{I}
\partial^{\mu} \Phi^{\bar {J}} + V\left(\Phi^{M},\Phi^{\bar M}\right) \right]
\end{equation}
and assuming that the universe is described by a FRW ansatz given by
\begin{equation}
ds^2 = -dt^2 + a(t)^2 d\Omega^2_{k}~,
\end{equation}
one can obtain the equations of motion for the moduli fields and the
metric 
\begin{equation}
{\ddot{\phi}}^{i}+3 \left(\frac{\dot a}{a}\right){\dot{\phi}}^{i}
+\Gamma^{i}_{jk}{\dot{\phi}}^{j}{\dot{\phi}}^{k}
+ {\sf G }^{ij} \frac{\partial V}{\partial\phi^{j}}=0~,
\end{equation}

\begin{equation}
\left(\frac{\dot{a}}{a}\right)^{2}+\frac{k}{a^{2}}
=\frac{1}{3}\left(\frac{1}{2} {\sf G}^{ij}{\dot{\phi}}^{i}{\dot{\phi}}^{j}
+ V\right)~.
\end{equation}
Note that $k=0,\pm 1$ parametrizes the spatial curvature of the $3d$ part of the
manifold, $\Gamma^{i}_{jk}$ are the Christoffel symbols
for the ${\sf G}^{ij}$  metric in field space and $\phi ^{i}$
denote the real components of the chiral fields such that
\begin{equation}
 K_{I \bar J} \partial_{\mu} \Phi^{I}
\partial^{\mu} \Phi^{\bar {J}} = \frac{1}{2}{\sf G}^{ij}\partial{\phi^{i}}\partial { \phi^{j}}~.
\end{equation}

Taking a single complex scalar field $\Phi = X +i Y$ we arrive at the
system of equations of the form,
\begin{eqnarray}
\label{XY-eom}
\ddot{X} &=&
-3\dot{X}\frac{\dot{a}}{a}+\frac{\dot{X^2}-\dot{Y^2}}{X}-\frac{2X^2V_{X}}{3}~,
\nonumber \\\nonumber\\
\ddot{Y} &=&
-3\dot{Y}\frac{\dot{a}}{a}+\frac{2\dot{X}\dot{Y}}{X}-\frac{2X^2V_{Y}}{3}~,
 \\ \nonumber  \\
\left(\frac
     {\dot{a}}{a}\right)^2 + \frac{k}{a^2} &=&
     \frac{\dot{X^2}+\dot{Y^2}}{4X^2} + \frac{V}{3}~. \nonumber
\end{eqnarray}

Once we have obtained the field trajectories
we can calculate the slow roll parameters at any point
using the general expressions for a 2 dimensional potential, 
\begin{equation}
\epsilon = \frac{1}{2} \left(\frac {{\sf G}^{ij} \partial_i V \partial_j V}{V^2} \right)~,
\end{equation} 
while $\eta$ is defined as the most-negative eigenvalue of
the matrix:
\begin {equation}
 N^i_j=\frac{{\sf G}^{ik} \left( \partial_k \partial_j V -
   \Gamma^l_{kj}\partial_l V \right)}{V}~.
\end{equation}

\subsection{Initial conditions for bubble universes.}

As we discussed in the main part of the text, we are interested 
in studying the evolution of the fields in the interior 
of a bubble universe that forms as a consequence of a
tunneling process. In order to do this, it is important 
to realize that the geometry of the bubble interior is
actually described by an infinite open universe \cite{CdL}. 
On the other hand, the Big Bang of this open universe
(the lightcone surface emanating from the nucleation center)
is perfectly smooth and could be thought of as a piece of
a Milne universe\footnote{Of course the whole history of the bubble
  interior is not describe by a Milne universe, only the early
  stages. Soon after the bubble formation the scale factor would evolve
  differently with time depending on the matter content
inside of the bubble.}. These constraints dictate that
the initial conditions for this type of geometry should be,
\begin{equation}
a(t) =  t + {\cal O}(t^3)
\end{equation}
and
\begin{eqnarray}
X(t) =  X^i_0 + {\cal O}(t^2)~,\\ \nonumber 
Y(t) =  Y^i_0 + {\cal O}(t^2)~, 
\end{eqnarray}
where $X^i_0$ and $Y^i_0$ are the ``exit point'' in field space
for the instanton that mediates between the parent and daughter 
vacua. Taking these initial conditions and the equations of
motion (\ref{XY-eom}) with $k=-1$, one can obtain the subsequent evolution
for the fields in the daughter bubble universe.

\section{Analytical Estimates of Probability distributions.}

We argued in the main part of the text that due to the nature of the 
attractor solution the results of our simulated landscape are quite
insensitive to the initial conditions for the fields. This allows the possibility of 
understanding the results in terms of a much simpler single field
inflation toy model\footnote{In this Appendix we follow closely the
discussion on \cite{D-brane-inflection-1} and \cite{D-brane-inflection-2}.}.

We start by assuming that all the realizations of our landscape can
locally be written around the inflection point as an expansion of the form
\begin{equation}
V \approx  V_0\left(1-\lambda_1 \phi - \lambda_3 \phi^3\right)~,
\end{equation}
where $\phi$ denotes the canonically normalized field and the parameters
of this expansion will be varying over our landscape. We can now
calculate all the observables in terms of these parameters in a
standard way. From the potential we get the slow roll inflation
parameters,
\begin{equation}
\label{epsilon}
\epsilon = \frac{1}{2} \left(\frac{V^\prime}{V}\right)^2 \approx
\frac{1}{2} \left(\lambda_1 +3\lambda_3 \phi^2 \right)^2 
\end{equation}
and
\begin{equation}
\label{eta}
\eta = \left(\frac{V^{\prime \prime}}{V}\right) \approx -6 \lambda_3 \phi
\end{equation}
as well as the total number of efolds
\begin{equation}
N_{total}(\lambda_1, \lambda_3) = \int_{-\infty}^\infty \frac{1}{\sqrt{2
    \epsilon}}{}\,\mathrm{d}\phi = \frac{\pi}{\sqrt{3\lambda_1\lambda_3}}~.
\end{equation}

From this relation we get the expression for the $\eta$ parameter at
the CMB scale, assuming that the end of inflation is given by $\eta\left(\phi_{end}\right)=-1$,
\begin{equation}
\eta_{CMB}(\lambda_1, \lambda_3)=\frac{2\pi}{N_{total}}\left[\tan\left[\frac{\pi
      N_{CMB}}{N_{total}}-\arctan\left[\frac{N_{total}}{2\pi}\right]\right]\right]~,
\end{equation}
which is function of only the first and the third derivative of
the potential. The spectral index can then be approximated by
\begin{equation}
\label{nsCMB}
n_s(\lambda_1, \lambda_3) = 1 + 2\eta_{CMB}\left(\lambda_1,\lambda_3\right)~.
\end{equation}

\begin{figure}[ht] 
\centering
\includegraphics[width=120mm]{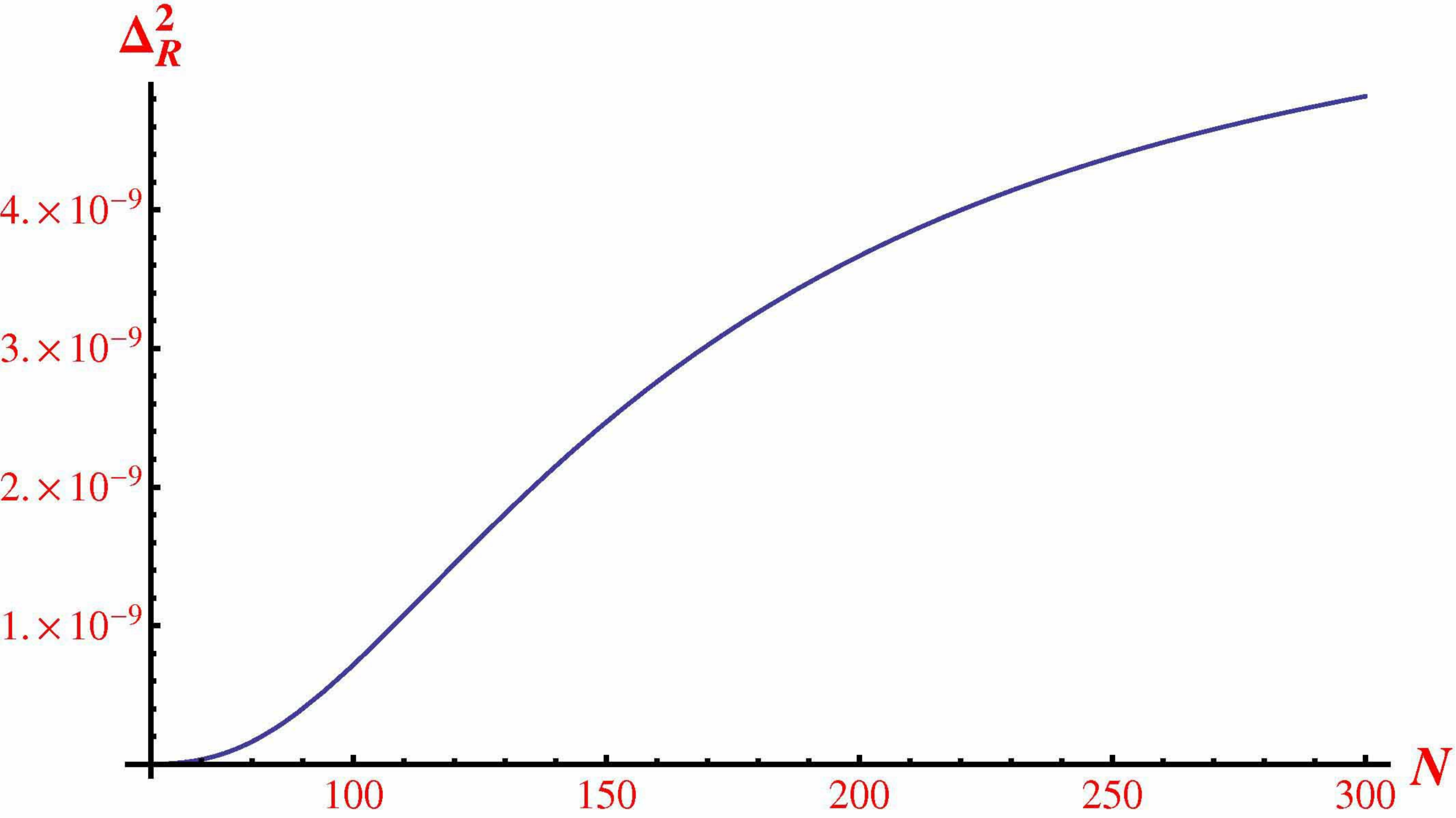}
\caption {Amplitude of perturbations versus the number of e-folds in a one dimensional landscape.}
\label{Amp-versus-N-A}
\end{figure}

Using the relations in Eqs.~(\ref{epsilon}) and (\ref{eta}) we
can approximate the position in field space $60$ e-folds before the
end of inflation as well as $\epsilon$ by the expressions
\begin{eqnarray}
\phi_{CMB} &=&-\frac{\eta_{CMB}}{6 \lambda_3}~, \\
\epsilon_{CMB} &=&
\frac{1}{2} \left(\lambda_1 +3\lambda_3 \phi_{CMB}^2 \right)^2~.
\end{eqnarray}
Using this information we can obtain the scalar power spectrum 
\begin{equation}
\label{deltaCMB}
\Delta_{\cal R}^2 (V_0,\lambda_1, \lambda_3) = \frac{1}{24\pi^2}\left.\frac{V}{\epsilon}\right|_{CMB}~.
\end{equation}

These equations will hold for any inflection point model
so they will apply for each of our realizations. We can then use
these relations to estimate the distribution of values over the
landscape assuming that one varies $W_0$ in a uniform way over 
the area that leads to more than $60$ e-folds.

Changing the superpotential induces small variations on the parameters of the inflection 
point potential which mostly do not change things significantly,
except the variation of $\lambda_1$. One can then model
the real landscape by assuming that one only scans uniformly over
this one parameter, $\lambda_1$. Taking into account that $N_{total}
\sim 1/ \sqrt{\lambda_1}$ one arrives at a distribution on the number
of e-folds of the form,
\begin{equation}
P(N) \sim {1\over {N^3}}~.
\end{equation}

\begin{figure}[ht] 
\centering
\includegraphics[width=120mm]{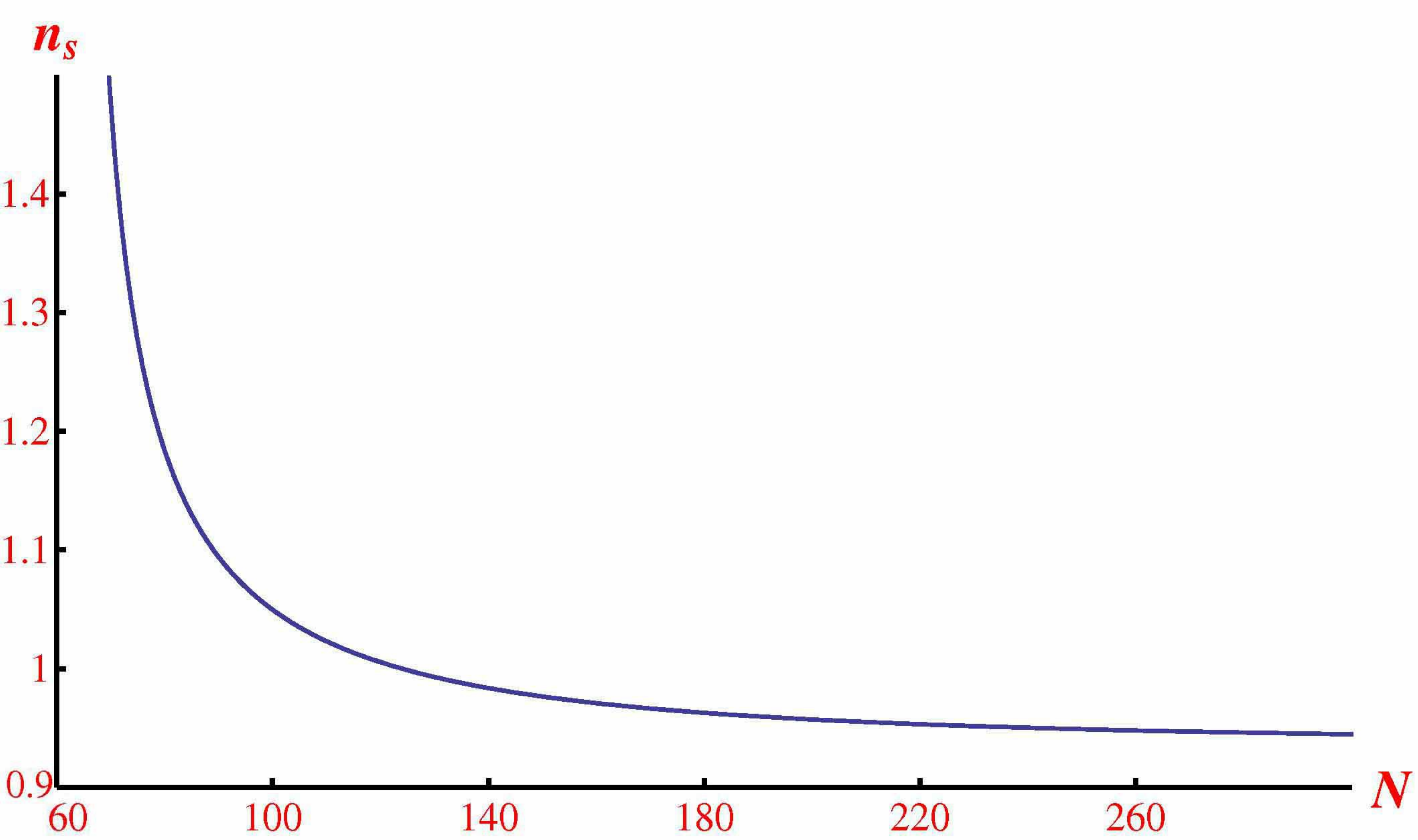}
\caption {$n_s$ versus the number of e-folds, $N$,  in a one dimensional landscape.} 
\label{ns-versus-N-A}
\end{figure}

Assuming this dependence of the number of e-folds with the variable
being scanned over the landscape, $\lambda_1$, and using the relations
found earlier in \ref{deltaCMB} and  \ref{nsCMB} one can find the
the amplitude of perturbations and the $n_s$ parameter
as a function of the number of e-folds. We plot these functions in 
Figs.~(\ref{Amp-versus-N-A}) and (\ref{ns-versus-N-A}).

Comparing these figures to the ones we found in our
random landscape we can infer that most of the simulated trajectories
closely follow this analytic form. This is due to the existence of the
attractor solution. There are however some special cases where
the trajectory never enters the attractor solution, but they are
statistically not very significant.

\end{document}